\documentclass[english]{article}
\usepackage[T1]{fontenc}
\usepackage[latin9]{inputenc}
\usepackage{color}
\usepackage{array}
\usepackage{rotating}
\usepackage{float}
\usepackage{units}
\usepackage{multirow}
\usepackage{amsthm}
\usepackage{amsmath}
\usepackage{amssymb}
\usepackage{wasysym}
\usepackage{graphicx}
\usepackage{esint}
\makeatletter

\providecommand{\tabularnewline}{\\}

\numberwithin{equation}{section}
\numberwithin{figure}{section}

\usepackage{babel}
\usepackage{fullpage}
\usepackage{authblk}
\usepackage{placeins}
\usepackage{graphics}
\title{Pricing and Hedging GLWB in the Heston and in the Black-Scholes
with Stochastic Interest Rate Models\footnote{This research was supported by  High-End Foreign Expert Project of Central University of Finance and Economics  (Beijing)}}
\author{ \textsc{Ludovic Goudenege}\thanks{F\'ederation de Math\'ematiques de l'\'Ecole Centrale Paris - CNRS FR3487 - \texttt{ ludovic.goudenege@math.cnrs.fr}} \and \textsc{Andrea Molent}\thanks{DEAMS, Universit\`a di Trieste - \texttt{andrea.molent@phd.units.it}} \and \textsc{Antonino Zanette}\thanks{Dipartimento di Scienze Economiche e Statistiche, Universit\`a di Udine - \texttt{antonino.zanette@uniud.it}}}
\date{}

\definecolor{blue}{rgb}{0,0.277,0.668}

\makeatother

\usepackage{babel}
\begin{document}
\maketitle

\begin{center}
\textbf{Abstract}
\par\end{center}

Valuing Guaranteed Lifelong Withdrawal Benefit (GLWB) has attracted
significant attention from both the academic field and real world
financial markets. As remarked by Forsyth and Vetzal \cite{FV} the
Black and Scholes framework seems to be inappropriate for such long
maturity products. They propose to use a regime switching model. Alternatively,
we propose here to use a stochastic volatility model (Heston model)
and a Black Scholes model with stochastic interest rate (Hull White
model). For this purpose we present four numerical methods for pricing
GLWB variables annuities: a hybrid tree-finite difference method and
a hybrid Monte Carlo method, an ADI finite difference scheme, and
a standard Monte Carlo method. These methods are used to determine
the no-arbitrage fee for the most popular versions of the GLWB contract,
and to calculate the Greeks used in hedging. Both constant withdrawal
and optimal withdrawal (including lapsation) strategies are considered.
Numerical results are presented which demonstrate the sensitivity
of the no-arbitrage fee to economic, contractual and longevity assumptions. 

\vspace{1cm}


\vspace{1cm}
\vspace{1cm}

\newpage

\section{Introduction}

In 2008 following the subprime crisis, financial markets have suffered
the upheavals that have affected the entire world economy. Since then,
these markets were extremely volatile: this situation could last a
while, and perhaps become the new standard. After many failures, the
gap between the different interest rate apply to different transmitters
has become larger and larger, and a discussion on the identification
of the risk-free rate is opened. The ECB and the Fed's rate gradually
declined, while rate on sovereign debt increased gradually.

For customers, it is difficult to balance risk and return. In this
context, clients seek protection for their savings, and the ability
to take advantage of the positive changes in the market. With regard
social problematic, following the increase in life expectancy, annuities
for retirement dropped. 

The mission of insurance companies is to answer the request for protection
and compensation of its customers. The solution is to provide the
customer an investment account and cover its value with guarantees.
These products are called Variable Annuities. In the words of Fran\c{c}ois
Robinet, CEO of AXA Life Invest, \textquotedbl{}These products, unit
of account guaranteed will become a solution to solve the long-term
investment problems with security, prepare for retirement\textquotedbl{}. 

In this article, we consider a Guaranteed Lifelong Withdrawal Benefit
(GLWB) annuity. We restrict our attention to a simplified form of
a GLWB which is initiated by making a lump sum payment to an insurance
company. This lump sum is then invested in risky assets, usually a
mutual fund. The benefit base, or guarantee account balance, is initially
set to the amount of the lump sum payment. The holder of the contract
is entitled to withdraw a fixed fraction of the benefit base for life,
even if the actual investment in the risky asset declines to zero.
Upon the death of the contract holder, his (her) estate receives the
remaining amount in the risky asset account. Typically, these contracts
have ratchet provisions (step-ups), that periodically increase the
benefit base if the risky asset investment has increased to a value
larger than the guarantee account value, and roll up provisions, that
periodically increase the benefit base according to a deterministic
function. In addition, the benefit base may also be increased if the
contract holder does not withdraw in a given year (bonus). Finally,
the contract holder may withdraw more than the contractually specified
amount, including complete surrender of the contract, upon payment
of a penalty. Complete surrender here means that the contract holder
withdraws the entire amount remaining in the investment account, and
the contract terminates. In most cases, this penalty for full or partial
surrender declines to zero after five to seven years.

The hedging costs for this guarantee are offset by deducting a proportional
fee from the risky asset account. From an insurance point of view,
these products are treated as financial ones: the products are hedged
as if they were pure financial products, and the mortality risk is
hedged using the law of large numbers. Therefore, it is very important
for insurance companies to be able to price quickly these products.
Moreover these products have long maturities that could last almost
60 years. The Black-Scholes model, with constant interest rate and
volatility seems to be unsuitable for those products: that\textquoteright{}s
why we present our pricing methods in two frameworks, modeling stochastic
volatility (Heston model \cite{He}) and stochastic interest rate
(Hull-White model \cite{HW}) . 

There have been several recent articles on pricing GLWBs. In particular,
we would remember the Forsyth and Vetzal's work \cite{FV}: they used
a PDE approach in a multi regimes model to price GLWBs contracts.
This approach proved to be very fast and accurate, and we used it
as a reference for our work. Concerning the use of stochastic volatility,
Kling et al. \cite{KL} used a Monte Carlo approach to price products.
We have made reference also to Bacinello et al. \cite{bips}: variable
annuities (including GLWBs) are priced using a Monte Carlo approach.
The policy holder (hereinafter, we will abbreviate it with \emph{PH})
behavior is assumed to be semi-Static, i.e. the holder withdraws at
the contract rate or surrenders the contract.

In this paper, we price GLWBs guarantees, and we find the no-arbitrage
fee, in the Heston model and the Black-Scholes with stochastic interest
rate model (\emph{BS HW model}). First, we treat a static withdrawal
strategy: the PH withdraws at the contract rate. Then, taking the
point of view of the worst case for the hedger, we price the guarantees
assuming that the contract holder follows an optimal withdrawal strategy.
We also used these methods to calculate the Greeks for hedging and
risk management. Moreover we performed a mortality shock useful in
risk management framework. For this purpose we present four numerical
methods: a hybrid tree-finite difference method and a hybrid Monte
Carlo method (both introduced by Briani et al. \cite{BCZ0}) an ADI
finite difference scheme (Haentjens and Hout \cite{HH}),  and a Standard
Monte Carlo method with Longstaff-Schwartz least squares regression
(Longstaff and Schwartz \cite{LS}). 

We use the term \emph{no-arbitrage fee} in the sense that this is
the fee which is required to maintain a replicating portfolio. A description
of the replicating portfolio for these types of guarantees is given
in Chen et al. \cite{CF} and Belanger et al. \cite{BF}. 

The main results of this paper are the following ones:
\begin{itemize}
\item We formulate the determination of the no-arbitrage fee (i.e. the cost
of maintaining a replicating hedging portfolio) in the Heston model
and in the BS HW model using different pricing methods;
\item We present the effects of stochastic volatility and stochastic interest
rate on pricing and Greeks calculation, and the sensitivity of the
GLWB fee to various modeling parameters;
\item We use different numerical methods to price the GLWB contract;
\item We present numerical examples which show the convergence of these
methods.
\end{itemize}
The paper is organized as follows: in Section 2, we describe the main
features of the contract such as mortality, withdrawals, and ratchets.
In Section 3, we provide a brief review of the stochastic models used
afterward. In Section 4, we present the numerical methods, and how
to implement them to solve the GLWB contract pricing problem. In Section
5 we perform tests in order to show their behavior and we study the
sensitivity of the no-arbitrage fee to economic, contractual and longevity
assumptions. Finally, in Section 6, we present the conclusions.

\section{\label{2}The GLWB Contract}

In the following, we will refer to the contract described in the paper
of Forsyth \cite{FV}, with some variations useful to compare our
results with other works. We make a brief summary of the main features
of the contract.

\subsection{Mortality}

We price the products in a risk-neutral measure, therefore in the
following we assume that mortality risk is diversifiable (Milevsky
and Salisbury, \cite{MS}). When this assumption is not justified,
then the risk-neutral value of the contract can be adjusted using
an actuarial premium principle (Gaillardetz and Lakhmiri, \cite{GL}).
Hereinafter, the time variable will be denoted by the letter $t$,
and we assume that the contract starts at $t=0$.

First we suppose that no PH can live longer than a given age. This
age will be denoted by $\tau$ (usually $\tau=122$). The age of the
PH at the beginning will be denoted by $a_{0}$ (usually $a_{0}=65$).
So, the maturity of the contract is $T=\tau-a_{0}$ (usually $T=57$):
when the time variable $t$ reaches $T$ all PHs are died, and the
contract is worth zero.

The effects of the mortality on the contract are described using two
functions:
\begin{itemize}
\item $\mathcal{M}:\left[0,T\right]\rightarrow\mathbb{R}$ is the probability
density that describes the random variable $M$ associated to the
death year of the PH. The fraction of the original owners who die
in $\left[t,t+dt\right]$ is equal to $\mathcal{M}\left(t\right)dt$.
\item $\mathcal{R}:\left[0,T\right]\rightarrow\mathbb{R}$ is the fraction
of the original owners who are still alive at time $t$
\[
\mathcal{R}\left(t\right)=1-\int_{0}^{t}\mathcal{M}\left(s\right)ds.
\]

\end{itemize}
We remark that $\mathcal{R}\left(0\right)=1$ and $\mathcal{R}\left(T\right)=0$.
For seek of simplicity, we assume $\mathcal{M}$ to be constant between
contract's anniversaries: if $t\in\left[k,k+1\right[$, $k\in\mathbb{N}$
then $\mathcal{M}\left(t\right)=\mathcal{M}\left(k\right)$.

\subsection{Contract State Parameters}

At time $t=0$ the policy holder pays with lump sum the gross premium
$GP$, to the insurance company. This may be reduced by some initial
fees, giving a net premium $P$. The premium $P$ is invested in a
fund, whose price is denoted by the variable $S_{t}$. The state parameters
of the contract are:
\begin{itemize}
\item Account value: $A_{t}$, $A_{0}=P$.
\item Base benefit: $B_{t}$, $B_{0}=GP$.
\end{itemize}
Both these two variables are initially set equal to the gross premium
or to the premium.

We suppose that the acquisition charges are equal to $GP-P$ aren't
used for hedging purposes, but only to cover entry costs for management
control. We suppose that there is a set of discrete times $t_{i}$,
which we term \emph{event times}. At these times, withdrawals, ratchets,
and bonuses may occur. Normally, event times are annually or quarterly.
We first consider the evolution of the value of the guarantee excluding
these event times $t_{i}$. 

The value of the contract at time $t$ is denoted by $V\left(A_{t},B_{t},t\right)$.

\subsection{Evolution of the Contract between Event Times.}

Let $t\in\left]t_{i},t_{i+1}\right[\subseteq\left[0,T\right]$. As
we said before, $S_{t}$ denotes the underlying fund driving the account
value. The dynamics of $S_{t}$ will be described in the next Section.
The account value $A_{t}$ follows the same dynamics of $S_{t}$ with
the exception of the fact that some fees may be subtracted continuously:

\begin{equation}
dA_{t}=\frac{A_{t}}{S_{t}}dS_{t}-\alpha_{tot}A_{t}dt.\label{eq:fees_cont}
\end{equation}
We suppose that total annual fees are charged to the policy holder
and withdrawn continuously from the investment account $A_{t}$. These
fees include the mutual fund management fees $\alpha_{m}$ and the
fee charged to fund the guarantee (also known as the rider) $\alpha_{g}$,
so that
\[
\alpha_{tot}=\alpha_{m}+\alpha_{g}.
\]
The only portion used by the insurance company to hedge the contract
is that coming from $\alpha_{g}$: the other fees has to be considered
as a outgoing money flow as PH's withdrawals are.

Continuously withdrawn fees are typical of the contract described
by Forsyth. Fees may also be withdrawn at the end of each policy year
$t_{i}$: this is what Kling et al. do in \cite{KL}. In this second
case

\begin{equation}
dA_{t}=\frac{A_{t}}{S_{t}}dS_{t}.\label{eq:fees_end}
\end{equation}
When the PH dies, a death benefit, usually equal to $A_{t}$, is payed
out to the heirs of the PH. According to the contract's formulation,
this death benefit may be payed immediately or at the upcoming event
time. If it is payed immediately, the contract stops immediately and
the account value and the benefit base becomes equal to zero; otherwise
the contract goes on up to the next event time as if nothing happened.

\subsection{Event Times}

An event time is a sequence of operations under the contract, which
occur at fixed dates, usually at each anniversary of the signing of
the contract. The times these events take place are denoted by $t_{i}=\Delta t\cdot i$
and usually $\Delta t=1$. Let's define $I=T/\Delta t$; then, $i$
runs in $\left\{ 0,\dots I\right\} $. 

When an event time occurs, we assume that the following events happen
in this order:
\begin{enumerate}
\item Withdrawal of the fees by the insurance company (if it is not time
continuous);
\item If the PH died, payment of the death benefits;
\item If the PH is still alive, he (she) is entitled to withdraw a certain
amount of money;
\item If provided by the contract, a ratchet may increase the benefit base
$B_{t}$.
\end{enumerate}
We denote with $\left(A_{t_{i}}^{-},B_{t_{i}}^{-},t_{i}\right)$ the
state variables just before an event time that occurs at time $t_{i}$
and with $\left(A_{t_{i}}^{k+},B_{t_{i}}^{k+},t_{i}\right)$ the state
variables just after the update due to the i-th point of the previous
numbered list.

\subsubsection{Fees}

Fees may be withdrawn continuously by the account value, as supposed
in Forsyth and Vetzal in\cite{FV}. In this case, between two event
times, the account value changes as prescribed by (\ref{eq:fees_cont}),
and nothing special happens at an event time:
\[
\left(A_{t_{i}}^{1+},B_{t_{i}}^{1+},t_{i}\right)=\left(A_{t_{i}}^{-},B_{t_{i}}^{-},t_{i}\right).
\]
 Otherwise, fees may be withdrawn at the end of the period, as supposed
in Kling et al. \cite{KL}. In this case, between two event times,
the account value changes as prescribed by (\ref{eq:fees_end}), and
at the event time, the account value becomes
\[
\left(A_{t_{i}}^{1+},B_{t_{i}}^{1+},t_{i}\right)=\left(A_{t_{i}}^{-}e^{-\alpha_{tot}\Delta t},B_{t_{i}}^{-},t_{i}\right).
\]
It is important to be able to deduce the management fees $F_{t}^{man}$
withdrawn by the account value because they are not used to hedge
the contract and therefore they have to be considered as an outgoing
money flow. If these fees are withdrawn continuously, we can calculate
them observing that their dynamic between two event times is 
\[
dF_{t}^{man}=\alpha_{m}A_{t}dt+r_{t}dt.
\]
This ODE has the following solution 
\[
F_{t}^{man}=\int_{0}^{t}e^{\int_{s}^{t}r_{u}du}\alpha_{m}A_{s}ds.
\]
and can be used in a Monte Carlo approach.

If the fees are withdrawn at the end of the period, we can calculate
management fees as a fraction of the total fees withdrawn: 
\[
F_{t_{i}}^{tot}=F_{t_{i-1}}^{tot}+A_{t_{i}}^{0}\left(1-e^{-\alpha_{tot}\Delta t}\right),
\]

\[
F_{t_{i}}^{man}=F_{t_{i-1}}^{man}+\frac{\alpha_{man}}{\alpha_{tot}}\left(F_{t_{i}}^{tot}-F_{t_{i-1}}^{tot}\right).
\]

\subsubsection{Death Benefit}

If the policy holder died at an instant $\bar{t}\in\left]t_{i-1},t_{i}\right[$
his (her) heirs will obtain a death benefit, that is usually equal
to the account value. If the contract provides that the death benefit
is paid immediately, then the death benefit $DB_{t}$ is paid in $t$
and is equal to $A_{t}$. Otherwise, if the DB is payed at the next
event time, $DB_{t_{i}}=A_{t_{i}}^{1+}$ and the contract is concluded
(after the DB payment it's worthless):
\[
\left(A_{t_{i}}^{2+},B_{t_{i}}^{2+},t_{i}\right)=\left(0,0,t_{i}\right).
\]

\subsubsection{\label{gamma}Withdrawal, Bonus, Surrender Event}

According to the contract, the policy holder, if still alive at event
time $t_{i}$, is entitled to withdraw a certain amount $W_{t_{i}}$
from his (her) police, also if the account value is equal to $0$.
This amount is given by 
\[
W_{t_{i}}=G\Delta t\cdot B_{t_{i}}^{2+},
\]
where $G$ is a constant defined by the contract. In a static framework,
we suppose that the PH simply withdraws $WA_{t_{i}}$. Otherwise,
in a optimization framework, he (she) may withdraw a fraction $\gamma_{i}$
of the guaranteed withdrawn:
\[
W_{t_{i}}=\gamma_{i}G\Delta t\cdot B_{t_{i}}^{2+}.
\]

\begin{itemize}
\item The case $\gamma_{i}=0$ corresponds to no withdrawal. In this case,
the contract may provide a bonus ($b_{t_{i}}$ is specified by the
contract):
\[
\left(A_{t_{i}}^{3+},B_{t_{i}}^{3+},t_{i}\right)=\left(A_{t_{i}}^{2+},B_{t_{i}}^{2+}\left(1+b_{t_{i}}\right),t_{i}\right).
\]

\item If $0<\gamma_{i}\leq1$ the PH withdraws at a lower rate than the
standard rate, and the new state variables are 
\[
\left(A_{t_{i}}^{3+},B_{t_{i}}^{3+},t_{i}\right)=\left(\max\left(0,A_{t_{i}}^{2+}-W_{t_{i}}\right),B_{t_{i}}^{2+},t_{i}\right).
\]

\item A third case is possible: the PH may want to withdraw more than the
maximum admitted. In this case we suppose $\gamma_{i}\in\left]1,2\right]$,
where the case $\gamma_{i}=2$ corresponds to a total surrender. We
define
\[
A'=\max\left(0,A_{t_{i}}^{2+}-G\Delta t\cdot B_{t_{i}}^{2+}\right).
\]
 The withdrawn amount is 
\[
W_{t_{i}}=G\Delta t\cdot B_{t_{i}}^{2+}+\left(\gamma_{i}-1\right)A'\left(1-\kappa_{t_{i}}\right).
\]
where $\kappa_{t_{i}}\in\left[0,1\right]$ is a penalty for withdrawal
above the contract amount. The new state variables are
\begin{align*}
\left(A_{t_{i}}^{3+},B_{t_{i}}^{3+},t_{i}\right) & =\left(\max\left(0,A_{t_{i}}^{2+}-G\Delta t\cdot B_{t_{i}}^{2+}-\left(\gamma_{i}-1\right)A'\right),\left(2-\gamma_{i}\right)B_{t_{i}}^{2+},t_{i}\right)\\
 & =\left(\left(2-\gamma_{i}\right)A',\left(2-\gamma_{i}\right)B_{t_{i}}^{2+},t_{i}\right).
\end{align*}

\end{itemize}

\subsubsection{Ratchet}

If the contract species a ratchet (step-up) feature, then the value
of the benefit base $B$ is increased if the investment account has
increased. The guarantee account $B$ can never decrease, unless the
contract is partially or fully surrendered:

\[
\left(A_{t_{i}}^{4+},B_{t_{i}}^{4+},t_{i}\right)=\left(A_{t_{i}}^{3+},\max\left(B_{t_{i}}^{3+},A_{t_{i}}^{3+}\right),t_{i}\right).
\]

Another feature that may be included in the contract is roll-up: for
seek of simplicity we won't treat this mechanism.

\subsection{Similarity Reduction}

An important property of GLWB contract is the fact that these contract
behave good under scaling transformations. If $\mathcal{V}\left(A,B,t\right)$
denotes the price of a contract, it is possible to prove that for
any scalar $\eta>0$
\begin{equation}
\eta\mathcal{V}\left(A,B,t\right)=\mathcal{V}\left(\eta A,\eta B,t\right).\label{eq:similarity}
\end{equation}

Then, we just have to treat the case $B=\hat{B}$ for a fixed $\hat{B}$
(for example $\hat{B}=P$), and then, choosing $\eta=\nicefrac{\hat{B}}{B}$,
we can obtain 
\[
\mathcal{V}\left(A,B,t\right)=\frac{B}{\hat{B}}\mathcal{V}\left(\frac{\hat{B}}{B}A,\hat{B},t\right),
\]

which means that we can solve the pricing problem only for a single
representative value of $B$. This effectively reduces the problem
dimension. The similarity reduction (\ref{eq:similarity}) was also
exploited from Shah et Bertsimas in \cite{SB}. We can observe how
the reduction similarity works both in the case of a contract that
does not contain mechanisms for increasing the base benefits (ratchet),
both for contracts with these properties.

\section{\label{3}The Stochastic Models of the Fund $S$}

To understand the different impacts of stochastic volatility and stochastic
interest rate over such a long maturity contract, we price the GLWB
VA according to two models: the Heston model, which provides stochastic
volatility, and the the Black-Scholes Hull-White model, which provide
stochastic interest rate. As we said before, the process $S$ represents
the underlying fund driving the product's account value $A_{t}$.

\subsection{The Heston Model}

The Heston model \cite{He} is one of the most known and used models
in finance to describe the evolution of the volatility of an underlying
asset and the underlying asset itself. In order to fix the notation,
we report its dynamics:
\begin{equation}
\begin{cases}
dS_{t}=rS_{t}dt+\sqrt{V_{t}}S_{t}dZ_{t}^{S} & S_{0}=\bar{S}_{0},\\
dV_{t}=k\left(\theta-V_{t}\right)dt+\omega\sqrt{V_{t}}dZ_{t}^{V} & V_{0}=\bar{V}_{0},
\end{cases}\label{eq:He}
\end{equation}
where $Z^{S}$ and $Z^{V}$ are Brownian motions, and $d\left\langle Z_{t}^{S},Z_{t}^{V}\right\rangle =\rho dt$.

\subsection{The Black-Scholes Hull-White Model}

The Hull-White model \cite{HW} is one of historically most important
interest rate models, which is nowadays often used for risk-management
purposes. The important advantage of the HW model is the existence
of the closed formulas for prices of bonds, caplets and swaptions.
In order to fix the notation, we report the dynamics of BS HW model:
\[
\begin{cases}
dS_{t}=r_{t}S_{t}dt+\sigma S_{t}dZ_{t}^{S} & S_{0}=\bar{S}_{0},\\
dr_{t}=k\left(\theta_{t}-r_{t}\right)dt+\omega dZ_{t}^{r} & r_{0}=\bar{r}_{0},
\end{cases}
\]
where $Z^{S}$ and $Z^{r}$ are Brownian motions, and $d\left\langle Z_{t}^{S},Z_{t}^{r}\right\rangle =\rho dt$. 

The process $r$ is a generalized Ornstein-Uhlenbeck (hereafter OU)
process: here $\theta_{t}$ is not constant but it is a deterministic
function which is completely determined by the market values of the
zero-coupon bonds by calibration (see Brigo and Mercurio \cite{BM}):
in this case the theoretical price of ZCB match exactly the market
prices.

Let $P^{M}\left(0,T\right)$ denote the market price of the zero bond
at time $0$ for the maturity $T$. The market instantaneous forward
interest rate is then defined by
\[
f^{M}\left(0,T\right)=-\frac{\partial\ln P^{M}\left(0,T\right)}{\partial T}.
\]
It is well known that the short rate process $r$ can be written as
\[
r_{t}=\omega X_{t}+\beta\left(t\right),
\]
where $X$  is a stochastic process given by 
\[
dX_{t}=-kX_{t}dt+dZ_{t}^{r},\ X_{0}=0,
\]
and $\beta\left(t\right)$ is a function
\[
\beta\left(t\right)=f^{M}\left(0,t\right)+\frac{\omega^{2}}{2k^{2}}\left(1-\exp\left(-kt\right)\right)^{2}.
\]
Then, the BS HW model is described by
\begin{equation}
\begin{cases}
dS_{t}=r_{t}S_{t}dt+\sigma S_{t}dZ_{t}^{S} & S_{0}=\bar{S}_{0},\\
dX_{t}=-kX_{t}dt+dZ_{t}^{r} & X_{0}=0,\\
r_{t}=\omega X_{t}+\beta\left(t\right).
\end{cases}\label{eq:HW}
\end{equation}

A particular case is called \emph{flat curve}. In this case, we assume
$P^{M}\left(t,T\right)=e^{-\bar{r}_{0}\left(T-t\right)}$ and $f^{M}\left(0,T\right)=\bar{r}_{0}$.
Then 
\[
\beta\left(t\right)=\bar{r}_{0}+\frac{\omega^{2}}{2k^{2}}\left(1-\exp\left(-kt\right)\right)^{2},
\]
 and

\[
\theta_{t}=\bar{r}_{0}+\frac{\omega^{2}}{2k^{2}}\left(1-\exp\left(-2kt\right)\right).
\]

\section{\label{4}Numerical Methods of Pricing}

In this Section we describe the four pricing methods: an Hybrid Monte
Carlo method, a standard Monte Carlo method, a Hybrid PDE method,
and an ADI PDE method.

We remember that our aim is to find the fair value for $\alpha_{g}$:
it's the value that makes the initial value of the policy equal to
the initial gross premium. To achieve this target, we price the police
(with one of the following procedures) and then we use the secant
method to approach the correct value for $\alpha_{g}$. Therefore,
the main goal is to be able to find the initial value for a given
value of $\alpha_{g}$: $V\left(A_{0},B_{0},0\right)\left(\alpha_{g}\right)$.

We remark that we want to calculate the value of the police from the
point of view of the insurance company: the management fees are treated
as a outgoing cash flows, and if we assume that the policy holder
follows a withdrawal strategy, we consider the worst one for the insurance
company.

\subsection{The Hybrid Monte Carlo Method}

The value of a GLWB police can be calculated through a Monte Carlo
set of simulations. This procedure is based in two steps: generation
of a scenario (a sampling of the underlying values along the life
of the product), and projection of the product into the scenario.
According to the way we obtain the scenarios, we distinguish two Monte
Carlo models: hybrid MC (HMC) and standard MC (SMC).

The hybrid MC method was introduced in Briani et al. \cite{BCZ}.
It is a simple and efficient way to produce MC scenarios for different
models. This method is called ``hybrid'' because it combines trees
and MC methods. First, a simple  tree needs to be built: this can
be done according to Appolloni et al. \cite{AC} and also \cite{NR},
or as we are going to show in \ref{tree}. Then, using a vector of
Bernoulli random variables, we move from the root through the tree,
describing the scenario for the volatility or the interest rate. The
values of the underlying at each time step can be easily obtained
using an Euler scheme.

\subsubsection{\label{tree}Trees}

\begin{figure}
\begin{centering}
\includegraphics[scale=0.7]{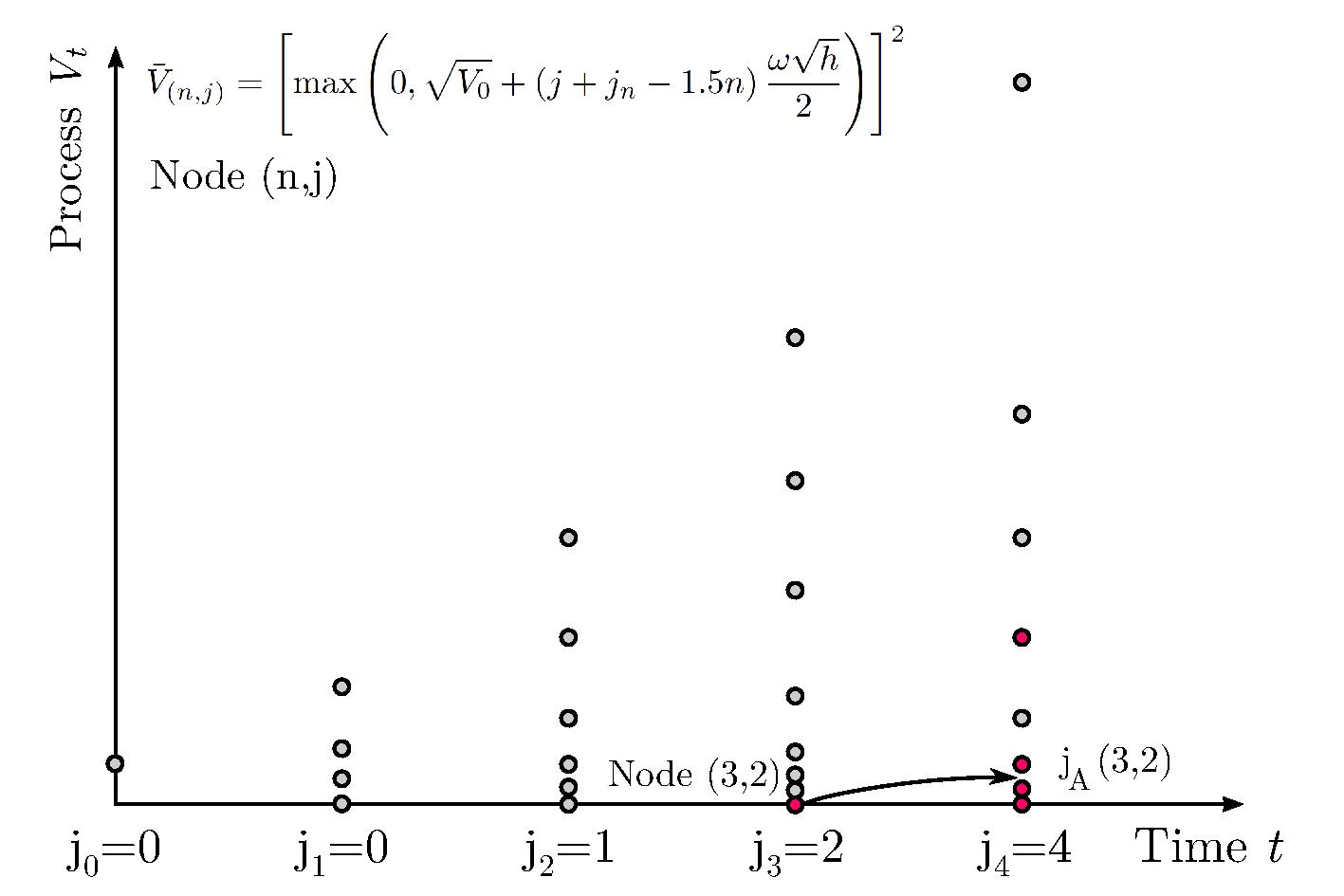}\includegraphics[scale=0.7]{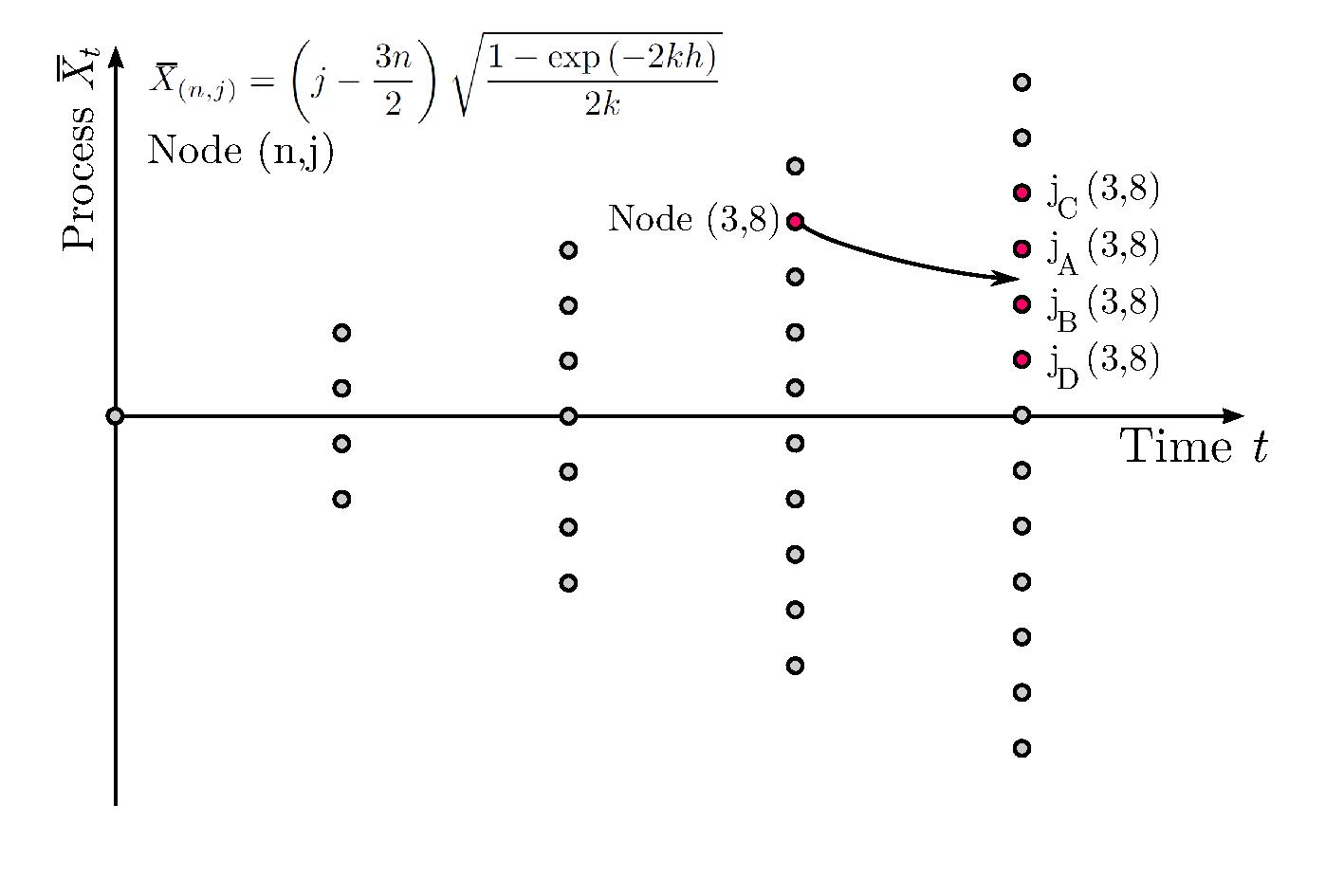}
\par\end{centering}

\caption{\label{fig:Trees}The trees for Heston and Hull-White models.}

\end{figure}

The trees for the Heston model and BS HW model can be obtained from
Appolloni et al. \cite{AC} or Nelson and Ramaswamy  \cite{NR}. In
this case, the trees are simple binary trees: the node values, and
the transition probabilities are set in order to match an approximation
of the first two moments of the processe. This kind of tree perform
well on short maturity, but the approximation errors accumulate on
long maturities. Because of this error that accumulates, the convergence
of the algorithm proved to be slow. Therefore, it was necessary to
rethink the trees: the main aim was to set up trees which matched
exactly some moments of the processes to be diffused. Here we present
two trees (see Figure \ref{fig:Trees}), one for stochastic volatility
and one for stochastic interest rate. They are simple quadrinomial
trees, and they are built to match the first 3 moments of the stochastic
processes.

We suppose to fix a number $N>0$, and we define $h=\nicefrac{T}{N}$.

\paragraph*{The General Case}

Let $Z$ be a Brownian motion, and let $G$ be a Gaussian stationary
process, following 
\[
dG_{t}=a\left(G_{t}\right)dt+bdZ_{t},
\]
 with variance that depends only by the time lapse, i.e. $G_{t+s}-G_{s}|\mathcal{F}_{s}\sim\mathcal{N}\left(\mu\left(t,G_{s}\right),\sigma^{2}\left(t\right)\right)$.
We show how to build a simple quadrinomial tree that can match the
first three moments.

We define a quadrinomial tree. Let's fix a maturity $T$, and the
number of steps $N$. Each node will be denoted by $G_{\left(n,j\right)}$
where $n$ runs from $0$ to $N$, and $j$ from $0$ to $3n$. Let
$h=\nicefrac{T}{N}$. The value of each node is 
\[
G_{\left(n,j\right)}=G_{0}+\left(j-1.5n\right)\sqrt{\sigma^{2}\left(h\right)}.
\]

We remember the first three moments of the process $G$: 
\[
M_{1}=\mathbb{E}\left[G_{t+h}-G_{t}|\mathcal{F}_{t}\right]=\mu\left(h,G_{t}\right),\ M_{2}=\mathbb{E}\left[\left(G_{t+h}-G_{t}\right)^{2}|\mathcal{F}_{t}\right]=\mu^{2}\left(h,G_{t}\right)+\sigma^{2}\left(h\right),
\]
\[
M_{3}=\mathbb{E}\left[\left(G_{t+h}-G_{t}\right)^{3}|\mathcal{F}_{t}\right]=\mu^{3}\left(h,G_{t}\right)+3\mu\left(h,G_{t}\right)\sigma^{2}\left(h\right).
\]

Let's fix a node $G_{\left(n,j\right)}$. To be brief, $\mu$ will
denote $\mu\left(h,G_{\left(n,j\right)}\right)$ and $\sigma$ will
denote $\sqrt{\sigma^{2}\left(h\right)}$. We suppose that the expected
value $\mu$ falls between the values of the nodes at time $\left(n+1\right)h$
. This hypothesis can be obtained assuming that the time step $h$
is small enough.

We define 
\[
j_{A}\left(n,j\right)=\mbox{ceil}\left[\frac{G_{0}-\mu}{\sigma}+1.5\left(n+1\right)\right],
\]

i.e. the first node in the next time step level whose value is bigger
than the mean of the process. This can be seen in Figure \ref{fig:Trees}
(both sides): the arrow points out to the expected value of the process,
and $j_{A}\left(n,j\right)$ is marked on the Figure. Let 
\[
j_{B}\left(n,j\right)=j_{A}\left(n,j\right)-1,\ j_{C}\left(n,j\right)=j_{A}\left(n,j\right)+1,\ j_{B}\left(n,j\right)=j_{D}\left(n,j\right)-2.
\]

To be brief we will only write $j_{A},j_{B},j_{C},$ $j_{D}$, and
$G_{A}$ will be $G_{A}=G_{\left(n+1,j_{A}\right)}$, and the same
for the other letters: this is clear in Figure \ref{fig:Trees}, on
the right side.

We can now define a Markovian discrete time process $\hat{G}_{n}$,
$n=0,\dots,N$ with $\hat{G}_{0}=G_{\left(0,0\right)}$ and we suppose
that if $\hat{G}_{n}=G_{\left(n,j\right)}$, then it can move to $G_{A}$,
$G_{B}$, $G_{C}$, $G_{D}$, according to the following probabilities
\[
p_{A}=P\left[\hat{G}_{n+1}=G_{A}|\hat{G}_{n}=G_{\left(n,j\right)}\right]=\frac{(G_{A}-\mu)\left((G_{A}-\sigma-\mu)^{2}+\sigma^{2}\right)}{2\sigma^{3}},
\]

\[
p_{B}=P\left[\hat{G}_{n+1}=G_{B}|\hat{G}_{n}=G_{\left(n,j\right)}\right]=\frac{(\mu-G_{A}+\sigma)\left((G_{A}-\mu)^{2}+\sigma^{2}\right)}{2\sigma^{3}},
\]

\[
p_{C}=P\left[\hat{G}_{n+1}=G_{C}|\hat{G}_{n}=G_{\left(n,j\right)}\right]=\frac{(\mu-G_{A}+\sigma)\left((G_{A}-\sigma-\mu)^{2}+2\sigma^{2}\right)}{6\sigma^{3}},
\]

\[
p_{D}=P\left[\hat{G}_{n+1}=G_{D}|\hat{G}_{n}=G_{\left(n,j\right)}\right]=\frac{2\sigma^{2}(G_{A}-\mu)+(G_{A}-\mu)^{3}}{6\sigma^{3}}.
\]

And since $G_{A}-\sigma<\mu\leq G_{A}$ , we can easily show that
these probabilities are well defined: all in $\left[0,1\right]$,
their sum is equal to $1$ , and the first three moments of the variable
$\hat{G}_{n+1}|\hat{G}_{n}=G_{\left(n,j\right)}$ are equals to the
first three moments of the variable $G_{t+h}|G_{t}=G_{\left(n,j\right)}$.

Now, we approximate the process $G$ by a discrete process $\bar{G}$
that is constant in each time lapse, and is defined as $\bar{G}_{t}=\hat{G}_{\left\lfloor \nicefrac{t}{N}\right\rfloor }$.
The weak convergence of this tree can be proved as in Nelson and Ramaswamy
 \cite{NR}.

\paragraph*{The Heston Model}

The Heston process (\ref{eq:He}) for volatility has no constant variance
and isn't Gaussian. We consider the process obtained by the square
root:
\[
d\sqrt{V_{t}}=\frac{4k\left(\theta-\sqrt{V_{t}}^{2}\right)-\omega^{2}}{8\sqrt{V_{t}}}dt+\frac{\omega}{2}dZ_{t}.
\]

We approximate it with a Gaussian process with variance $\frac{\omega^{2}}{4}dt$.
This approximation is helpful to define the grid: inspired by \cite{NR},
we define 
\[
j_{n}=\max\left(0,\mbox{floor}\left(1.5n-\frac{2\sqrt{V_{0}}}{\omega\sqrt{h}}\right)\right),
\]

and we set 
\[
\bar{V}_{\left(n,j\right)}=\left(\max\left(0,\sqrt{V_{0}}+\left(j+j_{n}-1.5n\right)\frac{\omega\sqrt{h}}{2}\right)\right)^{2}.
\]

for $j=0,\dots,3n-j_{n}$. The shift due to $j_{n}$ helps to reject
the many node with value equal to zero: if $j_{n}>0$, then $\bar{V}_{\left(n,0\right)}=0$
and $\bar{V}_{\left(n,1\right)}>0$ .

We fix now the value of $n$ and $j$. The discrete process $\bar{V}$
can jump from a node to another, as in a Markovian chain. We show
now how to find the possible upcoming nodes.

The first three moments for the Heston process can be found in Alfonsi
\cite{AA}:
\[
\psi\left(h\right)=\nicefrac{\left(1-e^{-kh}\right)}{k},\ \ \ M_{1}=\mathbb{E}\left[V_{t+h}|V_{t}=v\right]=ve^{-kh}+\theta k\psi\left(h\right),
\]

\[
M_{2}=\mathbb{E}\left[\left(V_{t+h}\right)^{2}|V_{t}=v\right]=M_{1}^{2}+\omega^{2}\psi\left(h\right)\left[\theta k\psi\left(h\right)/2+ve^{-kh}\right],
\]

\[
M_{3}=\mathbb{E}\left[\left(V_{t+h}\right)^{3}|V_{t}=v\right]=M_{1}M_{2}+\omega^{2}\psi\left(h\right)\left[2v^{2}e^{-2kh}+\psi\left(h\right)\left(k\theta+\frac{\omega^{2}}{2}\right)\left(3ve^{-kh}+\theta k\psi\left(h\right)\right)\right].
\]

Then we can proceed as in the general case. Anyway, the grid we're
using is based on an approximation: so the probabilities obtained
solving the linear system may not be positive.

\begin{figure}
\begin{centering}
\includegraphics{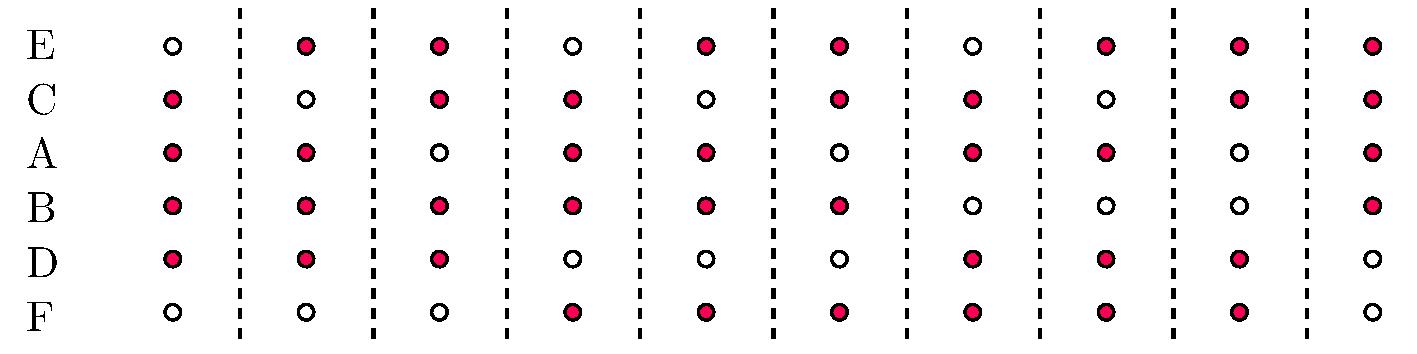}
\par\end{centering}

\caption{\label{fig:casi}The possible combinations used to get positive probabilities
in the Heston model tree. The red points correspond to the nodes used.}

\end{figure}

If we get negative probability for a given node, we try another combination
of nodes: the node $A$ or $C$ may be replaced by a node $E$ define
as the first node bigger than $C$, and the node $B$ or $D$ may
be replaced with a node $F$, defined as the smallest before node
$D$. This gives rise to $9$ combinations to be tested. If the starting
node is small and the node $D$ verifies $j_{D}=j_{n}$ we could not
do this last change because there would be no $F$ node. In this case
we allow the node $D$ to be replaced by the node $E$: see Figure
\ref{fig:casi}.

If these attempts don't give a positive result (negative probabilities),
we give up trying to match the first three moments, and we are content
to match an approximation of the first two as in \cite{NR}, thus
ensuring the weak convergence. In this case, we only use the nodes
$A,B,C,D$: we define 
\[
p_{AB}=\frac{\mu-G_{n+1,j_{B}}}{G_{n+1,j_{A}}-G_{n+1,j_{B}}},\ \ p_{BA}=1-p_{AB},
\]
\[
p_{CD}=\frac{\mu-G_{n+1,j_{D}}}{G_{n+1,j_{C}}-G_{n+1,j_{D}}},\ \ p_{DC}=1-p_{CD},
\]

and 
\[
p_{A}=\frac{5}{8}p_{AB},\ p_{B}=\frac{5}{8}p_{BA},\ p_{C}=\frac{3}{8}p_{CD},\ p_{D}=\frac{3}{8}p_{DC}.
\]

It is possible to show that the first moment of this variable is equal
to $M_{1}$, and as $h\rightarrow0$ the second moment approaches
to $M_{2}$, ensuring the convergence, as proved in \cite{NR}.

In all our numerical tests, this last option (matching only two moments)
has never been necessary: changing the nodes, all moments were matched
with positive probabilities.

\paragraph*{The Hull-White Model}

The process $X$ in (\ref{eq:HW}) is Gaussian. As shown in Ostrovski
\cite{VO} the variables $X_{t}$ and $\int_{s}^{t}X_{y}dy$ are bivariate
normal distributed conditionally on $X_{s}$ with well known mean
and variance. We define 
\[
X_{\left(n,j\right)}=\left(j-\frac{3n}{2}\right)\sqrt{\frac{1-\exp\left(-2kh\right)}{2k}},\ n=0,\dots,N\mbox{ and }j=0,\dots,3n.
\]

Let's fix a node $X_{\left(n,j\right)}$. We define
\[
H=\exp\left(-kh\right),\ K=\sqrt{\frac{1-\exp\left(-2kh\right)}{2k}},\ M_{1}=X{}_{\left(n,j\right)}H,
\]
 
\[
j_{A}=\mbox{ceil}\left[\frac{M_{1}}{K}+\frac{3\left(n+1\right)}{2}\right],\ X_{A}=X_{\left(n+1,j_{A}\right)}.
\]

The transition probabilities are given by
\[
\begin{array}{cc}
p_{A}=\frac{\left(X_{A}-M_{1}\right)}{2K^{3}}\left(K^{2}+\left(K+M_{1}-X_{A}\right)^{2}\right), & p_{B}=\frac{\left(K+M_{1}-X_{A}\right)}{2K^{3}}\left(K^{2}+\left(M_{1}-X_{A}\right)^{2}\right),\\
p_{C}=\frac{\left(K+M_{1}-X_{A}\right)}{6K^{3}}\left(2K^{2}+\left(K+M_{1}-X_{A}\right)^{2}\right), & p_{D}=\frac{\left(X_{A}-M_{1}\right)}{6K^{3}}\left(2K^{2}+\left(M_{1}-X_{A}\right)^{2}\right).
\end{array}
\]

\subsubsection{Scenario Generation}

The generations of the volatility process and of the interest rate
process behave in a similar way: we start from the node $\left(0,0\right)$
of the tree and according to a discrete random variable and to the
node probabilities, we move to the next node and so on. Let $D$ be
a discrete random variable that can assume value $A,B,C,D$ with probabilities
$p_{A},p_{B},p_{C},p_{D}$: sampling such a variable at each node,
we get the values of the process at each time step.

We distinguish two cases for the two models.

\paragraph*{The Heston Model}

We approximate the couple $\left(S_{t},V_{t}\right)$ in $\left[0,T\right]$
by a discrete process $\left(\bar{S}_{k\Delta t},\bar{V}_{k\Delta t}\right)_{k=0,\dots,T/\Delta t}$,
with $\left(\bar{S}_{0},\bar{V}_{0}\right)=\left(S_{0},V_{0}\right)$.
For each scenario, we generate the volatility. 

Let $N\sim\mathcal{N}\left(0,1\right)$ and $B\sim\mathcal{B}\left(0.5\right)$.
We deduce the value of $\bar{S}_{t+\Delta t}$ by 
\[
\bar{S}_{t+\Delta t}=\begin{cases}
\bar{S}_{t}\exp\!\left[\!\left(r-\frac{\rho}{\sigma}k\theta\right)\!\Delta t\!+\!\left(\frac{\rho}{\sigma}k-\frac{1}{2}\right)\!\!\left(\!\frac{\bar{V}_{t+\Delta t}+\bar{V}_{t}}{2}\!\right)\!\Delta t\!+\!\frac{\rho}{\sigma}\!\left(\bar{V}_{t+\Delta t}-\bar{V}_{t}\right)\!+\!\sqrt{\!\left(1-\rho^{2}\right)\Delta t\bar{V}_{t}}N\!\right] & \mbox{if }B=0,\\
\bar{S}_{t}\exp\!\left[\!\left(r-\frac{\rho}{\sigma}k\theta\right)\!\Delta t\!+\!\left(\frac{\rho}{\sigma}k-\frac{1}{2}\right)\!\!\left(\!\frac{\bar{V}_{t+\Delta t}+\bar{V}_{t}}{2}\!\right)\!\Delta t\!+\!\frac{\rho}{\sigma}\!\left(\bar{V}_{t+\Delta t}-\bar{V}_{t}\right)\!+\!\sqrt{\!\left(1-\rho^{2}\right)\Delta t\bar{V}_{t+\Delta t}}N\!\right] & \mbox{if }B=1.
\end{cases}
\]

According to (\ref{eq:He}), we use the normal variable $N$ to generate
the Gaussian increment of $S$, and the Bernoulli variable $B$ to
split the operator associated to the Heston process.

This scheme (without splitting) appears in Briani et al. \cite{BCZ}
and the splitting method appears in Alfonsi \cite{AA}.

\paragraph*{The Black-Scholes Hull-White Model}

We approximate the couple $\left(S_{t},X_{t}\right)$ in $\left[0,T\right]$
by a discrete process $\left(\bar{S}_{k\Delta t},\bar{X}_{k\Delta t}\right)_{k=0,\dots,T/\Delta t}$,
with $\left(\bar{S}_{0},\bar{X}_{0}\right)=\left(S_{0},0\right)$,
ad we deduce the interest rate by $\bar{r}_{t}=\omega\bar{X}_{t}+\beta\left(t\right)$.
Let $N\sim\mathcal{N}\left(0,1\right)$. We deduce the value of $\bar{S}_{t+\Delta t}$
by
\[
\bar{S}_{t+\Delta t}=\bar{S}_{t}\exp\left[\left(\frac{\bar{r}_{t\Delta t}+\bar{r}_{t}}{2}-\frac{\sigma^{2}}{2}\right)\Delta t+\sigma\left(\left(\bar{X}_{t+\Delta t}+\bar{X}_{t}\left(k\Delta t-1\right)\right)\rho+\sqrt{\Delta t}\bar{\rho}N\right)\right].
\]

\subsubsection{\label{Proj}Projection}

Once we have generated the scenarios, we project the police into it:
it means we calculate the initial value of the contract as the sum
of discounted cash flows. This calculation depends on whether we take
an optimized strategy or not. Let $\mathcal{V}\left(A,B,t\right)$
be the value of a police at time $t$, having account value equal
to $A$ and base benefit equal to $B$. From now on, we fix a specific
scenario. Let $V\left(A,B,t\right)$be the value of a police in that
scenario, at time $t$, having account value equal to $A$ and base
benefit equal to $B$.

\paragraph*{Constant Withdrawal}

In this case the strategy of the PH is fixed: in each event time $\gamma_{i}=1$
(for completeness we continue to write $\gamma_{i}$). A simple way
to calculate the value of the police is calculating forward the cash
flows, conditioning on the death time. As in Holz et al.  \cite{HK},
we have:
\[
V\left(A_{0},B_{0},0\right)=\sum_{i=0}^{I}\mathcal{M}\left(t_{i}\right)\left(\sum_{k=0}^{i}e^{-\int_{0}^{t_{k}}r_{s}ds}W_{t_{k}}+e^{-\int_{0}^{t_{i}}r_{s}ds}A_{t_{i}}^{1+}\right).
\]

Anyway, we developed another approach, useful for the optimal withdrawal
case. First we calculate the values $\left(A_{t_{i}}^{4+},B_{t_{i}}^{4+},t_{i}\right)$
for all $t_{i}$ neglecting the effect of mortality (equivalently,
assuming that the PH die at the end), with a forward approach:

\[
A_{t_{i}}^{4+}=\max\left(0,A_{t_{i-1}}^{4+}\frac{S_{t_{i}}}{S_{t_{i-1}}}e^{-\alpha_{tot}\Delta t}-\gamma_{t_{i}}G\Delta tB_{t_{i-1}}^{4+}\right),
\]
\[
B_{t_{i}}^{4+}=\begin{cases}
\max\left(B_{t_{i-1}}^{4+},A_{t_{i}}^{4+}\right) & \mbox{if ratchet},\\
B_{t_{i-1}}^{2+} & \mbox{otherwise.}
\end{cases}
\]
Then, we proceed backwards, calculating the value of the contract
for each time $t_{i}$ just before the withdrawal. The value of the
contract at time $t_{i}$ can be written as the discounted value at
time $t_{i+1}$ plus the discounted value of the cash flows relating
the period $\left[t_{i}^{4+},t_{i+1}^{4+}\right]$. The final condition
on the value of the contract is 
\[
V\left(A_{T}^{4+},B_{T}^{4+},T\right)=0,
\]
 because all PHs are death and all benefits have been paid. Then
\[
V\left(A_{t_{i}}^{4+},B_{t_{i}}^{4+},t_{i}\right)=e^{-\int_{t_{i}}^{t_{i+1}}r_{s}ds}\left[V\left(A_{t_{i+1}}^{4+},B_{t_{i+1}}^{4+},t_{i+1}\right)+\mathcal{R}\left(t_{i+1}\right)W_{t_{i+1}}\right]+DB+MF,
\]
where DB and MF stand for the discounted value in $t_{i}$ of the
death benefit and management fees paid in $\left[t_{i}^{4+},t_{i+1}^{4+}\right]$.
We distinguish four cases depending on how the management fees and
the death benefit are payed.

\subparagraph*{CASE 1: DB payed at the end, Fees withdrawn at the end}

{\small{}
\[
\begin{array}{ccc}
DB=\mathcal{M}\left(t_{i}\right)e^{-\int_{t_{i}}^{t_{i+1}}r_{s}ds}A_{t_{i}}^{4+}\frac{S_{t_{i+1}}}{S_{t_{i}}}e^{-\alpha_{tot}\Delta t}, & \ \  & MF=\mathcal{R}\left(t_{i}\right)e^{-\int_{t_{i}}^{t_{i+1}}r_{s}ds}A_{t_{i}}^{4+}\frac{S_{t_{i+1}}}{S_{t_{i}}}\left(1-e^{-\alpha_{tot}\Delta t}\right)\frac{\alpha_{m}}{\alpha_{tot}}.\end{array}
\]
}{\small \par}

\subparagraph*{CASE 2: DB payed at the end, Fees withdrawn continuously}

{\small{}
\[
\begin{array}{ccc}
DB=\mathcal{M}\left(t_{i}\right)e^{-\int_{t_{i}}^{t_{i+1}}r_{s}ds}A_{t_{i}}^{4+}\frac{S_{t_{i+1}}}{S_{t_{i}}}e^{-\alpha_{tot}\Delta t}, & \ \  & MF=\mathcal{R}\left(t_{i}\right)\alpha_{m}\frac{A_{t_{i}}^{4+}}{S_{t_{i}}}\int_{t_{i}}^{t_{i+1}}e^{-\int_{t_{i}}^{t}r_{s}ds}S_{t}e^{-\alpha_{tot}\left(t-t_{i}\right)}dt.\end{array}
\]
}{\small \par}

\subparagraph*{CASE 3: DB payed immediately, Fees withdrawn at the end }

{\small{}
\[
DB=\mathcal{M}\left(t_{i}\right)\frac{A_{t_{i}}^{4+}}{S_{t_{i}}}\int_{t_{i}}^{t_{i+1}}e^{-\int_{t_{i}}^{t}r_{s}ds}S_{t}e^{-\alpha_{tot}\left(t-t_{i}\right)}dt,
\]
}{\small \par}

{\small{}
\[
MF=\mathcal{M}\left(t_{i}\right)\frac{\alpha_{m}}{\alpha_{tot}}\frac{A_{t_{i}}^{4+}}{S_{t_{i}}}\int_{t_{i}}^{t_{i+1}}S_{t}\left(1-e^{-\alpha_{tot}\left(t-t_{i}\right)}\right)e^{-\int_{t_{i}}^{t}r_{u}du}dt+\mathcal{R}\left(t_{i+1}\right)e^{-\int_{t_{i}}^{t_{i+1}}r_{s}ds}A_{t_{i}}^{4+}\frac{S_{t_{i+1}}}{S_{t_{i}}}\left(1-e^{-\alpha_{tot}\Delta t}\right)\frac{\alpha_{m}}{\alpha_{tot}}.
\]
}{\small \par}

\subparagraph*{CASE 4: DB payed immediately, Fees withdrawn continuously}

{\small{}
\[
DB=\mathcal{M}\left(t_{i}\right)\frac{A_{t_{i}}^{4+}}{S_{t_{i}}}\int_{t_{i}}^{t_{i+1}}e^{-\int_{t_{i}}^{t}r_{s}ds}S_{t}e^{-\alpha_{tot}\left(t-t_{i}\right)}dt,
\]
}{\small \par}

{\small{}
\[
MF=\mathcal{M}\left(t_{i}\right)\alpha_{m}\frac{A_{t_{i}}^{4+}}{S_{t_{i}}}\int_{t_{i}}^{t_{i+1}}S_{t}e^{-\alpha_{tot}\left(t-t_{i}\right)}e^{-\int_{t_{i}}^{t}r_{u}du}\left(t_{i+1}-t\right)dt+\mathcal{R}\left(t_{i+1}\right)\alpha_{m}\frac{A_{t_{i}}^{4+}}{S_{t_{i}}}\int_{t_{i}}^{t_{i+1}}e^{-\int_{t_{i}}^{t}r_{s}ds}S_{t}e^{-\alpha_{tot}\left(t-t_{i}\right)}dt.
\]
}Proceeding in this way, it is possible to calculate $V\left(A_{0}^{4+},B_{0}^{4+},0\right)$.
The initial value of the police is
\[
V\left(A_{0}^{-},B_{0}^{-},0\right)=V\left(A_{0}^{4+},B_{0}^{4+},0\right),
\]
if the first withdrawal takes place at time $t=t_{1}$, or 
\[
V\left(A_{0}^{-},B_{0}^{-},0\right)=V\left(A_{0}^{4+},B_{0}^{4+},0\right)+\gamma_{0}G\Delta tP
\]
if the first withdrawal takes place at time $t=0$. Then we simply
have to calculate the average of $V\left(A_{0}^{-},B_{0}^{-},0\right)$
among the simulated scenarios.

\paragraph*{Optimal Withdrawal}

In this case we suppose that at each event time $t_{i}$ the policy
holder can withdraw a fraction $\gamma_{i}$ of the regular amount.
To price in this case, we suppose that the PH chooses the value of
$\gamma$ that causes the worst hedging case for the insurance company.
We denote $\mathcal{V}\left(A,B,t\right)$ the expected value at time
$t$ of a generic police whose state parameters are $A,B$ :
\[
\mathcal{V}\left(A,B,t\right)=\mathbb{E}\left[V\left(A,B,t\right)\right].
\]
So, we suppose that the policy holder chooses $\gamma_{i}$ such that
\[
\gamma_{i}=\underset{\gamma\in\left[0,2\right]}{\mbox{argmax}}\ \mathcal{V}\left(A^{4+},B^{4+},t\right).
\]
This expected value can be calculated with a Longstaff-Schwartz approach:
\begin{enumerate}
\item Simulate $N$ random scenarios and price the police into these scenarios
using random values for $\gamma_{i}$.
\item For $t=T$ to $t=0$:

\begin{enumerate}
\item Approximate the function $\mathcal{V}\left(A,B,t\right)$ using the
least squares projection into a space of functions (usually polynomials).
\item For each scenario find the optimal value of $\gamma_{t}$ .
\item Recalculate the upcoming state variables from $s=t$ to $s=T$ assuming
that the policy holder chooses the best value for $\gamma$.
\end{enumerate}
\item Calculate the average of the initial value $V\left(A_{0},B_{0},0\right)$
for all the scenarios.
\end{enumerate}
The approximation of the function $\mathcal{V}\left(A,B,t\right)$
can be improved by the reduction property.

\subsection{Standard Monte Carlo Method \label{MC}}

The Monte Carlo method is very similar to the hybrid Monte Carlo one.
The only different thing, is the way we produce the random scenarios.
The projection phase is the same as hybrid Monte Carlo's one.

\subsubsection{Scenario generation}

We distinguish two cases for the two models.

\paragraph*{The Heston Model}

The generation of the scenarios (underlying and volatility) in this
case has been done using a third order schemes described in Alfonsi
\cite{AA}.

\paragraph*{The Black-Scholes Hull-White Model}

The generation of the scenarios (underlying and interest rate) in
this case has been done using an exact schemes described in Ostrovski
\cite{VO}, with a few changes in order to incorporate the correlation
between underlying and interest rate.

\subsection{PDE Hybrid Method}

The Hybrid PDE approach is different from the previous ones. In fact
it's a PDE pricing method and it's based on Briani et al. \cite{BCZ0},\cite{BCZ}
both for Heston and Hull-White case. Using a tree to diffuse volatility
or interest rate, we freeze these values between two tree-levels and
we solve four PDE for each tree node; then we mix the values given
by the four PDEs according to the transition probabilities of each
node.

We can resume the pricing methods in three features: model, algorithm
structure and pricing.

\subsubsection{The Heston Model }

Starting from the model
\[
\begin{cases}
dS_{t}=rS_{t}dt+\sqrt{V_{t}}S_{t}\left(\rho dZ_{t}^{V}+\bar{\rho}dZ_{t}^{S}\right) & V_{0}=\bar{V}_{0},\\
dV_{t}=k\left(\theta-V_{t}\right)dt+\omega\sqrt{V_{t}}dZ_{t}^{V} & S_{0}=\bar{S}_{0},
\end{cases}\ \ \ d\left\langle Z_{t}^{S},Z_{t}^{V}\right\rangle =0,
\]
we define the process 
\[
E_{t}=\ln\left(A_{t}\right)-\frac{\rho}{\omega}V_{t},\ E_{0}=\ln\left(A_{0}\right)-\frac{\rho}{\omega}V_{0},
\]
\begin{equation}
A_{t}=\exp\left(E_{t}+\frac{\rho}{\omega}V_{t}\right).\label{eq:invE}
\end{equation}
Then 
\[
dE_{t}=\left(r-\frac{V_{t}}{2}-\frac{\rho}{\omega}k\left(\theta-V_{t}\right)-\alpha_{tot}\right)dt+\sqrt{\left(1-\rho^{2}\right)V_{t}}dZ_{t}^{S},
\]
if fees are taken continuously, otherwise
\[
dE_{t}=\left(r-\frac{V_{t}}{2}-\frac{\rho}{\omega}k\left(\theta-V_{t}\right)\right)dt+\bar{\rho}\sqrt{\left(1-\rho^{2}\right)V_{t}}dZ_{t}^{S}.
\]

\subsubsection{The Black-Scholes Hull-White Model}

Starting from the model

\[
\begin{cases}
dS_{t}=r_{t}S_{t}dt+\sigma S_{t}\left(\rho dZ_{t}^{r}+\bar{\rho}dZ_{t}^{S}\right) & S_{0}=\bar{S}_{0},\\
dX_{t}=-kX_{t}dt+dZ_{t}^{r} & X_{0}=0,\\
r_{t}=\omega X_{t}+\beta\left(t\right),
\end{cases}\ \ \ d\left\langle Z_{t}^{S},Z_{t}^{r}\right\rangle =0,
\]
we define the process 
\[
U_{t}=\ln\left(A_{t}\right)-\rho\sigma X_{t},\ U_{0}=\ln\left(A_{0}\right),
\]
\begin{equation}
A_{t}=\exp\left(U_{t}+\rho\sigma X_{t}\right).\label{eq:invU}
\end{equation}
Then 
\[
dU_{t}=\left(r_{t}-\frac{\sigma^{2}}{2}+\sigma\rho kX_{t}-\alpha_{tot}\right)dt+\sigma\sqrt{1-\rho^{2}}dZ_{t}^{S},
\]
if fees are taken continuously, otherwise
\[
dU_{t}=\left(r_{t}-\frac{\sigma^{2}}{2}+\sigma\rho kX_{t}\right)dt+\sigma\sqrt{1-\rho^{2}}dZ_{t}^{S}.
\]

\subsubsection{Algorithm structure}

The structures for this algorithm consist in a tree and a PDE solver.
As described in Briani et al. \cite{BCZ0},\cite{BCZ}, we use a tree
to diffuse the volatility (or the interest rate) along the life of
the product, and we solve backward a 1D PDE freezing at each node
of the tree the volatility (or the interest rate). The tree is built
according to Section \ref{tree} (quadrinomial tree, matching the
first three moments of the process), and the PDE is solved with a
finite difference approach. We have to solve the PDE between event
time, and at the event time we apply the changes to the states to
reproduce the effects of the events.

\subsubsection{\label{Pric}Pricing}

The PDE we have to solve is essentially the same as in Forsyth and
Vetzal \cite{FV}. We distinguish four cases as we did in Monte Carlo
case. We denote with $\mathcal{V}\left(A,B,t\right)$ the value of
a contract at time $t$, whose account value is worth $A$ and whose
base benefit is worth $B$ . Consequently, we define 
\[
\mathcal{V}^{He}\left(E,B,t\right)=\mathcal{V}\left(\exp\left(E+\frac{\rho}{\omega}V_{t}\right),B,t\right),
\]
and 
\[
\mathcal{V}^{HW}\left(U,B,t\right)=\mathcal{V}\left(\exp\left(U+\rho\sigma X_{t}\right)B,t\right).
\]
The variables $\bar{r}$, $\bar{X}$ and $\bar{V}$ will denote the
frozen values of $r_{t}$, $X_{t}$ and $V_{t}$. We solve the transformed
PDE between two event times for each node of the tree four times:
one for each of the possible next nodes, using the initial data corresponding
to these nodes. To reduce the run time, we do this only for most relevant
nodes: this cutting technique dramatically reduced calculation times
without compromising the quality of results. Then, using the inverse
transformations (\ref{eq:invE}) and (\ref{eq:invU}), we apply the
event times' actions. In the next few paragraphs, we are going to
write 2 PDEs: one for the Heston model, and one for the BS HW model.

\subsubsection*{CASE 1: DB payed at the end, Fees withdrawn at the end}

The terminal condition is 
\[
\mathcal{V}\left(A,B,T\right)=\mathcal{R}\left(T-\Delta t\right)A\left(1-\left(1-e^{-\alpha_{tot}\Delta t}\right)\frac{\alpha_{g}}{\alpha_{tot}}\right).
\]
 The associated PDEs are 

{\small{}
\[
\mathcal{V}_{t}^{He}+\frac{\bar{\rho}^{2}\bar{V}}{2}\mathcal{V}_{EE}^{He}+\left(r-\frac{\bar{V}}{2}-\frac{\rho}{\omega}k\left(\theta-\bar{V}\right)\right)\mathcal{V}_{E}^{He}-r\mathcal{V}^{He}=0\ \tag{He 1},
\]
}{\small \par}

{\small{}
\[
\mathcal{V}_{t}^{HW}+\frac{\bar{\rho}^{2}\sigma^{2}}{2}\mathcal{V}_{UU}^{HW}+\left(\bar{r}-\frac{\sigma^{2}}{2}+\sigma\rho k\bar{X}\right)\mathcal{V}_{U}^{HW}-\bar{r}\mathcal{V}^{HW}=0\ \tag{HW 1}.
\]
}For $t_{i}=T-1$ to $t_{i}=0$ we have to:
\begin{enumerate}
\item Solve the PDE backward from $t_{i+1}$ to $t_{i}$.
\item Calculate the value of $\mathcal{V}$ from the value of $\mathcal{V}^{He}$
or $\mathcal{V}^{HW}$.
\item In case of ratchet $\mathcal{V}\left(A,B,t_{i}^{3+}\right)=\mathcal{V}\left(A,\max\left(A,B\right),t_{i}^{4+}\right)$.
\item Withdrawal: 

\begin{enumerate}
\item if $\gamma_{t_{i}}=0$ :{\small{}
\[
\mathcal{V}\left(A,B,t_{i}^{2+}\right)=\mathcal{V}\left(A,B\left(1+b_{t_{i}}\right),t_{i}^{3+}\right);
\]
}{\small \par}
\item if $\gamma_{t_{i}}\in\left[0,1\right]$ :{\small{}
\[
\mathcal{V}\left(A,B,t_{i}^{2+}\right)=\mathcal{V}\left(\max\left(0,A-\gamma_{t_{i}}G\Delta tB\right),B,t_{i}^{3+}\right)+\mathcal{R}\left(t_{i}\right)\gamma_{t_{i}}G\Delta tB;
\]
}{\small \par}
\item if $\gamma_{t_{i}}\in\left]1,2\right]$ :{\small{}
\begin{multline*}
\mathcal{V}\left(A,B,t_{i}^{2+}\right)=\mathcal{V}\left(\max\left(0,A-G\Delta tB\right)\left(2-\gamma_{t_{i}}\right),B\left(2-\gamma_{t_{i}}\right),t_{i}^{3+}\right)+\\
+\mathcal{R}\left(t_{i}\right)\left(G\Delta tB+\left(\gamma_{t_{i}}-1\right)\max\left(0,A-G\Delta tB\right)\left(1-\kappa_{t_{i}}\right)\right);
\end{multline*}
}{\small \par}
\end{enumerate}
\item Death benefit: $\mathcal{V}\left(A,B,t_{i}^{1+}\right)=\mathcal{V}\left(A,B,t_{i}^{2+}\right)+\left(\mathcal{R}\left(t_{i-1}\right)-\mathcal{R}\left(t_{i}\right)\right)A$.
\item Fees: $\mathcal{V}\left(A,B,t_{i}^{-}\right)=\mathcal{V}\left(Ae^{-\alpha_{tot}\Delta t},B,t_{i}^{1+}\right)+\mathcal{R}\left(t_{i-1}\right)\frac{\alpha_{m}}{\alpha_{tot}}A\left(1-e^{-\alpha_{tot}\Delta t}\right)$.
\item Calculate the value of $\mathcal{V}^{He}$ or $\mathcal{V}^{HW}$
from the value of $\mathcal{V}$ .
\end{enumerate}

\subsubsection*{CASE 2: DB payed at the end, Fees withdrawn continuously}

The differences between this case and the case 1 are the following
ones. The terminal condition is 
\[
\mathcal{V}\left(A,B,T\right)=\mathcal{R}\left(T-\Delta t\right)A.
\]
The associated PDEs are 

{\small{}
\[
\mathcal{V}_{t}^{He}+\frac{\bar{\rho}^{2}\bar{V}}{2}\mathcal{V}_{EE}^{He}+\left(r-\frac{\bar{V}}{2}-\frac{\rho}{\omega}k\left(\theta-\bar{V}\right)-\alpha_{tot}\right)\mathcal{V}_{E}^{He}-r\mathcal{V}^{He}+\alpha_{m}\mathcal{R}\left(t\right)\exp\left(E_{t}+\frac{\rho}{\omega}\bar{V}\right)=0\ \tag{He 2},
\]
}{\small \par}

{\small{}
\[
\mathcal{V}_{t}^{HW}+\frac{\bar{\rho}^{2}\sigma^{2}}{2}\mathcal{V}_{UU}^{HW}+\left(\bar{r}-\frac{\sigma^{2}}{2}+\sigma\rho k\bar{X}-\alpha_{tot}\right)\mathcal{V}_{U}^{HW}-\bar{r}\mathcal{V}^{HW}+\alpha_{m}\mathcal{R}\left(t\right)\exp\left(U_{t}+\rho\sigma\bar{X}\right)=0\ \tag{HW 2}.
\]
}Point 6 (fees step) becomes 
\[
\mathcal{V}\left(A,B,t_{i}^{-}\right)=\mathcal{V}\left(A,B,t_{i}^{1+}\right).
\]

\subsubsection*{CASE 3: DB payed immediately, Fees withdrawn at the end }

The differences between this case and the case 1 are the following
ones. The terminal condition is 
\[
\mathcal{V}\left(A,B,T\right)=0.
\]
The associated PDEs are 

{\small{}
\[
\mathcal{V}_{t}^{He}\!+\!\frac{\bar{\rho}^{2}\bar{V}}{2}\mathcal{V}_{EE}^{He}\!+\!\left(\! r\!-\!\frac{\bar{V}}{2}\!-\!\frac{\rho}{\omega}k\left(\theta-\bar{V}\right)\!\right)\!\mathcal{V}_{E}^{He}\!-\! r\mathcal{V}^{He}\!+\!\mathcal{M}\left(t_{i}\right)\exp\left(E_{t}\!+\!\frac{\rho}{\omega}\bar{V}\right)\left(\!1\!-\!\left(\!1\!-\! e^{-\alpha_{tot}\left(t-t_{i}\right)}\!\right)\!\frac{\alpha_{g}}{\alpha_{tot}}\!\right)=0\ \tag{He 3},
\]
}{\small \par}

{\small{}
\[
\mathcal{V}_{t}^{HW}\!+\!\frac{\bar{\rho}^{2}\sigma^{2}}{2}\mathcal{V}_{UU}^{HW}\!+\!\left(\!\bar{r}\!-\!\frac{\sigma^{2}}{2}\!+\!\sigma\rho k\bar{X}\!\right)\mathcal{V}_{U}^{HW}\!-\!\bar{r}\mathcal{V}^{HW}\!+\!\mathcal{M}\left(t_{i}\right)\exp\left(U_{t}\!+\!\rho\sigma\bar{X}\right)\!\left(\!1\!-\!\left(\!1\!-\! e^{-\alpha_{tot}\left(t-t_{i}\right)}\!\right)\frac{\alpha_{g}}{\alpha_{tot}}\!\right)=0\ \tag{HW 3}.
\]
}Point 5 (death benefit step) and 6 (fees step) become:
\begin{itemize}
\item Death benefit: $\mathcal{V}\left(A,B,t_{i}^{1+}\right)=\mathcal{V}\left(A,B,t_{i}^{2+}\right)$.
\item Fees: $\mathcal{V}\left(A,B,t_{i}^{-}\right)=\mathcal{V}\left(Ae^{-\alpha_{tot}\Delta t},B,t_{i}^{1+}\right)+\mathcal{R}\left(t_{i}\right)\frac{\alpha_{m}}{\alpha_{tot}}A\left(1-e^{-\alpha_{tot}\Delta t}\right).$
\end{itemize}

\subsubsection*{CASE 4: DB payed immediately, Fees withdrawn continuously}

The differences between this case and the case 1 are the following
ones. The terminal condition is 
\[
\mathcal{V}\left(A,B,T\right)=0.
\]
The associated PDEs are 

{\small{}
\[
\mathcal{V}_{t}^{He}+\frac{\bar{\rho}^{2}\bar{V}}{2}\mathcal{V}_{EE}^{He}+\left(r-\frac{\bar{V}}{2}-\frac{\rho}{\omega}k\left(\theta-\bar{V}\right)-\alpha_{tot}\right)\mathcal{V}_{E}^{He}-r\mathcal{V}^{He}+\exp\left(E_{t}+\frac{\rho}{\omega}\bar{V}\right)\left(\alpha_{m}\mathcal{R}\left(t\right)+\mathcal{M}\left(t_{i}\right)\right)=0\ \tag{He 4},
\]
}{\small \par}

{\small{}
\[
\mathcal{V}_{t}^{HW}+\frac{\bar{\rho}^{2}\sigma^{2}}{2}\mathcal{V}_{UU}^{HW}+\left(\bar{r}-\frac{\sigma^{2}}{2}+\sigma\rho k\bar{X}-\alpha_{tot}\right)\mathcal{V}_{U}^{HW}-\bar{r}\mathcal{V}^{HW}+\exp\left(U_{t}+\rho\sigma\bar{X}\right)\left(\alpha_{m}\mathcal{R}\left(t\right)+\mathcal{M}\left(t_{i}\right)\right)=0\ \tag{HW 4}.
\]
}Point 5 (death benefit step) and 6 (fees step) become
\[
\mathcal{V}\left(A,B,t_{i}^{-}\right)=\mathcal{V}\left(A,B,t_{i}^{1+}\right)=\mathcal{V}\left(A,B,t_{i}^{2+}\right).
\]

\medskip

This concludes the static withdrawal case. In the optimal withdrawal
case, we suppose the PH to change the value of $\gamma_{i}$ used
in step n. 4 (withdrawal step). He (she) will choose the value of
$\gamma_{i}\in\left[0,2\right]$ in order to maximizes the value of
$\mathcal{V}\left(A,B,t_{i}^{2+}\right)$. This maximization can be
done using a grid of values for $\gamma_{i}$ and choosing at each
time the best value.

\subsection{PDE ADI Method}

Consider the asset price process given by the system of stochastic
 differential equations described in Section \ref{2}.  We describe
the ADI method only in the case 2, but the other cases can be easily
adapted. Moreover, we have chosen to not use the transformed PDE described
in Section \ref{Pric}, but the classical version of PDEs for the
Black-Scholes, Heston and Black-Scholes Hull-White model. The associated
PDEs are

{\small{} \[ \mathcal{V}_{t}^{He}+\frac{VA^{2}}{2}\mathcal{V}_{AA}^{He} + \frac{\omega^{2}V}{2}\mathcal{V}_{VV}^{He} +\left(r-\alpha_{tot}\right)A\mathcal{V}_{A}^{He}+\rho \omega AV \mathcal{V}_{AV}^{He}+k \left(\theta -V\right)\mathcal{V}_{V}^{He}-r\mathcal{V}^{He}+\alpha_{m}\mathcal{R}\left(t\right)A=0\ \tag{He 2b} \] }{\small \par}

{\small{} \[ \mathcal{V}_{t}^{HW}+\frac{\sigma^{2}A^{2}}{2}\mathcal{V}_{AA}^{HW} + \frac{\omega^{2}}{2}\mathcal{V}_{rr}^{HW} +\left(r-\alpha_{tot}\right)A\mathcal{V}_{A}^{HW}+\rho \omega A\sigma \mathcal{V}_{Ar}^{HW}+k \left(\theta_{t} -r\right)\mathcal{V}_{r}^{HW}-r\mathcal{V}^{HW}+\alpha_{m}\mathcal{R}\left(t\right)A=0\ \tag{HW 2b} \] }{\small \par}
         
Because of the long maturity, solving a two-dimensional PDE is a very costly and slow method. The idea is to use splitting schemes of ADI (alternating directional  implicit) type. In this paper, we only present the Douglas scheme, but  various scheme are available in the literature.
In order to solve the PDE, we should address many numerical difficulties.
The first one is the mesh and we have chosen to use the  meshes described in \cite{HH} with the parameters \[ A_{left} = 0.8 S_{0}\quad A_{right}=1.2S_{0} \quad A_{max}=100 S_{0}  \quad \text{ and } d_{1} = S_{0}/20, \] for the mesh of variable $A$,  \[ R_{\max} = 10 R_{0}, \quad c=R_{0}\quad \text{ and } d_{2} = R_{\max}/400 \] for the mesh of variable $r$ in the Black-Scholes Hull-White model, and \[ V_{\max} = MIN(MAX(100V_{0},1),5) \quad \text{ and } d_{3} = V_{\max}/500. \] for the mesh of variable $V$ in the Heston model.
The second difficulty is the choice of the splitting scheme. We have  chosen the Douglas scheme with parameter $\theta =1/2$ because it is the  easiest to implement, but of course some higher order schemes (in time)  would be more optimal.
The last difficulty, but not the least, is the choice of boundary  conditions. Since there is no closed form solutions for the GLWB  product, it is difficult to make the right choice for the boundary  conditions. Moreover the boundary conditions have a big impact on the  solution, because of the long maturity. 
The choice of  homogeneous Neumann conditions is usually done because it simplifies the  system to solve (exactly it simplifies the finite difference scheme at  the boundary). In the context of GLWB, the boundary conditions for the Black-Scholes Hull-White model will be given by: \begin{eqnarray*} \frac{\partial \mathcal{V}_{t}^{HW}}{\partial s} (A,r,t)= 0, & &\text{ if } A=0 \text{ or }  A=A_{\max},\\  \frac{\partial \mathcal{V}_{t}^{HW}}{\partial v} (A,r,t)= 0, & &\text{ if } r=\pm R_{\max}, \end{eqnarray*} on the mesh $[0,A_{\max}]\times [-R_{\max},R_{\max}]$, and the boundary condition for the Heston model will be given by:  \begin{eqnarray*} \frac{\partial \mathcal{V}_{t}^{He}}{\partial s} (A,V,t)= 0, & &\text{ if } A=0 \text{ or }  A=A_{\max},\\  \frac{\partial \mathcal{V}_{t}^{He}}{\partial v} (A,V,t)= 0, & &\text{ if } V=V_{\max}, \end{eqnarray*}  on the mesh $[0,A_{\max}]\times [0,V_{\max}]$,  and with no condition at $V=0$ since it is an outflow boundary.

\section{\label{5}Numerical Results}

In this Section we compare the numerical methods used in Section \ref{4}:
Hybrid Monte Carlo (\emph{HMC}), Standard Monte Carlo (\emph{SMC}),
Hybrid PDE (\emph{HPDE}), and ADI PDE (\emph{APDE}). In particular
we compare pricing and Greeks computation in \emph{Static Case} \ref{5.1}
and \emph{Dynamic Case} \ref{5.2}.

We chose the parameters of the methods according to 4 configurations
(\emph{A, B, C, D}), with an increasing number of steps, and so that
the calculation time for the various methods in each configuration
were close. There 4 configurations are in Table \ref{tab:Configuration-parameters},
with the notation (time steps per year ; space steps; vol steps) for
the ADI PDE method, (time steps per year ; space steps ) for the Hybrid
PDE method approaches, and (time steps per year ; number of simulations)
for the MC's one. In Monte Carlo for dynamic case, we also add the
degree of the approximating polynomial. These values had been chosen
to achieve approximately these run times: $\left(A\right)$ 30 s,
$\left(B\right)$ 120 s, $\left(C\right)$ 480 s, $\left(D\right)$
1900 s. To reduce the run time we do the secant iterations using an
increasing number of time steps for all the methods: the values in
Table \ref{tab:Configuration-parameters} are those used for the last
3 iterations.

We use the standard MC both as a pricing method, both as a benchmark
(BM). About the benchmark, in the static case we used $10^{7}$ independent
runs. In the dynamic case we used $10^{6}$ independent runs, arranged
in $10$ sub runs; in each sub runs the expected value has been approximated
by a $6$ order polynomial. At each event time, the PH can chose between
$\gamma=0$, $\gamma=1$ and $\gamma=2$. 

The search for the fair $\alpha_{g}$ value has been driven by the
secant method. The initial values for this method were $\alpha_{g}=0$
bp and $\alpha_{g}=200$ bp.

To achieve Delta calculation in Monte Carlo methods we used a $1\permil$
shock in static case and $1\%$ in dynamic case.

We used the DAV 2004R mortality Table, 65 year old German male (see
\cite{FV} for the Table). It contains the probabilities that a person
aged $t$ will die within the next year. It's easy to get the function
$\mathcal{M}$ from these probabilities.

\begin{table}
\begin{centering}
{\footnotesize{}}%
\begin{tabular}{|>{\centering}p{0.15cm}|>{\centering}p{1.6cm}>{\centering}p{1.6cm}>{\centering}p{1.4cm}>{\centering}p{1.6cm}||>{\centering}p{1.6cm}>{\centering}p{1.6cm}>{\centering}p{1.4cm}>{\centering}p{1.7cm}|}
\cline{2-9} 
\multicolumn{1}{>{\centering}p{0.15cm}|}{} & \multicolumn{4}{c||}{\textsc{\small{}BS HW Static}} & \multicolumn{4}{c|}{\textsc{\small{}Heston Static}}\tabularnewline
\multicolumn{1}{>{\centering}p{0.15cm}|}{} & \textsc{\footnotesize{}HMC} & \textsc{\footnotesize{}SMC} & \textsc{\footnotesize{}HPDE} & \textsc{\footnotesize{}APDE} & \textsc{\footnotesize{}HMC} & \textsc{\footnotesize{}SMC} & \textsc{\footnotesize{}HPDE} & \textsc{\footnotesize{}APDE}\tabularnewline
\hline 
\emph{\footnotesize{}A} & \textcolor{blue}{\footnotesize{}$5\negthinspace\times\negthinspace1.3\negthinspace\cdot\negthinspace10^{5}$} & \textcolor{blue}{\footnotesize{}$1\negthinspace\times\negthinspace2.7\negthinspace\cdot\negthinspace10^{5}$} & \textcolor{blue}{\footnotesize{}$30\negthinspace\times\negthinspace400$} & \textcolor{blue}{\footnotesize{}$18\negthinspace\times\negthinspace180\negthinspace\times\negthinspace36$} & \textcolor{blue}{\footnotesize{}$5\negthinspace\times\negthinspace8.6\negthinspace\cdot\negthinspace10^{4}$} & \textcolor{blue}{\footnotesize{}$5\negthinspace\times\negthinspace7.4\negthinspace\cdot\negthinspace10^{4}$} & \textcolor{blue}{\footnotesize{}$35\negthinspace\times\negthinspace400$} & \textcolor{blue}{\footnotesize{}$26\negthinspace\times\negthinspace260\negthinspace\times\negthinspace13$}\tabularnewline
\emph{\footnotesize{}B} & \textcolor{blue}{\footnotesize{}$10\negthinspace\times\negthinspace2.3\negthinspace\cdot\negthinspace10^{5}$} & \textcolor{blue}{\footnotesize{}$1\negthinspace\times\negthinspace9.8\negthinspace\cdot\negthinspace10^{5}$} & \textcolor{blue}{\footnotesize{}$60\negthinspace\times\negthinspace600$} & \textcolor{blue}{\footnotesize{}$27\negthinspace\times\negthinspace270\negthinspace\times\negthinspace54$} & \textcolor{blue}{\footnotesize{}$10\negthinspace\times\negthinspace1.6\negthinspace\cdot\negthinspace10^{5}$} & \textcolor{blue}{\footnotesize{}$10\negthinspace\times\negthinspace1.4\negthinspace\cdot\negthinspace10^{5}$} & \textcolor{blue}{\footnotesize{}$70\negthinspace\times\negthinspace600$} & \textcolor{blue}{\footnotesize{}$40\negthinspace\times\negthinspace400\negthinspace\times\negthinspace20$}\tabularnewline
\emph{\footnotesize{}C} & \textcolor{blue}{\footnotesize{}$20\negthinspace\times\negthinspace5.4\negthinspace\cdot\negthinspace10^{5}$} & \textcolor{blue}{\footnotesize{}$1\negthinspace\times\negthinspace4.9\negthinspace\cdot\negthinspace10^{6}$} & \textcolor{blue}{\footnotesize{}$100\negthinspace\times\negthinspace1000$} & \textcolor{blue}{\footnotesize{}$40\negthinspace\times\negthinspace400\negthinspace\times\negthinspace80$} & \textcolor{blue}{\footnotesize{}$20\negthinspace\times\negthinspace3.8\negthinspace\cdot\negthinspace10^{5}$} & \textcolor{blue}{\footnotesize{}$20\negthinspace\times\negthinspace3.5\negthinspace\cdot\negthinspace10^{5}$} & \textcolor{blue}{\footnotesize{}$100\negthinspace\times\negthinspace1000$} & \textcolor{blue}{\footnotesize{}$64\negthinspace\times\negthinspace640\negthinspace\times\negthinspace32$}\tabularnewline
\emph{\footnotesize{}D} & \textcolor{blue}{\footnotesize{}$40\negthinspace\times\negthinspace1.0\negthinspace\cdot\negthinspace10^{6}$} & \textcolor{blue}{\footnotesize{}$1\negthinspace\times\negthinspace2.0\negthinspace\cdot\negthinspace10^{7}$} & \textcolor{blue}{\footnotesize{}$200\negthinspace\times\negthinspace2000$} & \textcolor{blue}{\footnotesize{}$62\negthinspace\times\negthinspace620\negthinspace\times\negthinspace124$} & \textcolor{blue}{\footnotesize{}$40\negthinspace\times\negthinspace7.3\negthinspace\cdot\negthinspace10^{5}$} & \textcolor{blue}{\footnotesize{}$40\negthinspace\times\negthinspace7.5\negthinspace\cdot\negthinspace10^{5}$} & \textcolor{blue}{\footnotesize{}$200\negthinspace\times\negthinspace2000$} & \textcolor{blue}{\footnotesize{}$104\negthinspace\times\negthinspace1040\negthinspace\times\negthinspace52$}\tabularnewline
\hline 
\end{tabular}
\par\end{centering}{\footnotesize \par}

{\footnotesize{}\vspace{0.1cm}}{\footnotesize \par}

\begin{centering}
{\footnotesize{}}%
\begin{tabular}{|>{\centering}p{0.15cm}|>{\centering}p{1.6cm}>{\centering}p{1.6cm}>{\centering}p{1.4cm}>{\centering}p{1.6cm}||>{\centering}p{1.6cm}>{\centering}p{1.6cm}>{\centering}p{1.4cm}>{\centering}p{1.7cm}|}
\cline{2-9} 
\multicolumn{1}{>{\centering}p{0.15cm}|}{} & \multicolumn{4}{c||}{\textsc{\small{}BS HW Dynamic}} & \multicolumn{4}{c|}{\textsc{\small{}Heston Dynamic}}\tabularnewline
\multicolumn{1}{>{\centering}p{0.15cm}|}{} & \textsc{\footnotesize{}HMC} & \textsc{\footnotesize{}SMC} & \textsc{\footnotesize{}HPDE} & \textsc{\footnotesize{}APDE} & \textsc{\footnotesize{}HMC} & \textsc{\footnotesize{}SMC} & \textsc{\footnotesize{}HPDE} & \textsc{\footnotesize{}APDE}\tabularnewline
\hline 
\emph{\footnotesize{}A} & \textcolor{blue}{\footnotesize{}$5\negthinspace\times\negthinspace3.3\!\cdot\!10^{3}\negthinspace\times\negthinspace2$} & \textcolor{blue}{\footnotesize{}$5\negthinspace\times\negthinspace3.2\!\cdot\!10^{3}\negthinspace\times\negthinspace2$} & \textcolor{blue}{\footnotesize{}$30\negthinspace\times\negthinspace400$} & \textcolor{blue}{\footnotesize{}$16\negthinspace\times\negthinspace160\negthinspace\times\negthinspace32$} & \textcolor{blue}{\footnotesize{}$5\negthinspace\times\negthinspace3.2\!\cdot\!10^{3}\negthinspace\times\negthinspace2$} & \textcolor{blue}{\footnotesize{}$5\negthinspace\times\negthinspace3.2\!\cdot\!10^{3}\negthinspace\times\negthinspace2$} & \textcolor{blue}{\footnotesize{}$35\negthinspace\times\negthinspace400$} & \textcolor{blue}{\footnotesize{}$22\negthinspace\times\negthinspace220\negthinspace\times\negthinspace11$}\tabularnewline
\emph{\footnotesize{}B} & \textcolor{blue}{\footnotesize{}$10\negthinspace\times\negthinspace1.6\!\cdot\!10^{4}\negthinspace\times\negthinspace3$} & \textcolor{blue}{\footnotesize{}$5\negthinspace\times\negthinspace1.6\!\cdot\!10^{4}\negthinspace\times\negthinspace3$} & \textcolor{blue}{\footnotesize{}$60\negthinspace\times\negthinspace600$} & \textcolor{blue}{\footnotesize{}$24\negthinspace\times\negthinspace240\negthinspace\times\negthinspace48$} & \textcolor{blue}{\footnotesize{}$10\negthinspace\times\negthinspace1.5\!\cdot\!10^{4}\negthinspace\times\negthinspace3$} & \textcolor{blue}{\footnotesize{}$10\negthinspace\times\negthinspace1.5\!\cdot\!10^{4}\negthinspace\times\negthinspace3$} & \textcolor{blue}{\footnotesize{}$70\negthinspace\times\negthinspace600$} & \textcolor{blue}{\footnotesize{}$36\negthinspace\times\negthinspace360\negthinspace\times\negthinspace18$}\tabularnewline
\emph{\footnotesize{}C} & \textcolor{blue}{\footnotesize{}$20\negthinspace\times\negthinspace5.2\!\cdot\!10^{4}\negthinspace\times\negthinspace4$} & \textcolor{blue}{\footnotesize{}$5\negthinspace\times\negthinspace5.3\!\cdot\!10^{4}\negthinspace\times\negthinspace4$} & \textcolor{blue}{\footnotesize{}$100\negthinspace\times\negthinspace1000$} & \textcolor{blue}{\footnotesize{}$38\negthinspace\times\negthinspace380\negthinspace\times\negthinspace76$} & \textcolor{blue}{\footnotesize{}$20\negthinspace\times\negthinspace4.9\!\cdot\!10^{4}\negthinspace\times\negthinspace4$} & \textcolor{blue}{\footnotesize{}$20\negthinspace\times\negthinspace4.9\!\cdot\!10^{4}\negthinspace\times\negthinspace4$} & \textcolor{blue}{\footnotesize{}$100\negthinspace\times\negthinspace1000$} & \textcolor{blue}{\footnotesize{}$60\negthinspace\times\negthinspace600\negthinspace\times\negthinspace30$}\tabularnewline
\emph{\footnotesize{}D} & \textcolor{blue}{\footnotesize{}$40\negthinspace\times\negthinspace1.4\!\cdot\!10^{5}\negthinspace\times\negthinspace5$} & \textcolor{blue}{\footnotesize{}$5\negthinspace\times\negthinspace1.6\!\cdot\!10^{5}\negthinspace\times\negthinspace5$} & \textcolor{blue}{\footnotesize{}$200\negthinspace\times\negthinspace2000$} & \textcolor{blue}{\footnotesize{}$60\negthinspace\times\negthinspace600\negthinspace\times\negthinspace120$} & \textcolor{blue}{\footnotesize{}$40\negthinspace\times\negthinspace1.3\!\cdot\!10^{5}\negthinspace\times\negthinspace5$} & \textcolor{blue}{\footnotesize{}$40\negthinspace\times\negthinspace1.3\!\cdot\!10^{5}\negthinspace\times\negthinspace5$} & \textcolor{blue}{\footnotesize{}$200\negthinspace\times\negthinspace2000$} & \textcolor{blue}{\footnotesize{}$100\negthinspace\times\negthinspace1000\negthinspace\times\negthinspace50$}\tabularnewline
\hline 
\end{tabular}
\par\end{centering}{\footnotesize \par}

\caption{\label{tab:Configuration-parameters}Configuration parameters for
the BS HW model and for the Heston model, static and dynamic.}

\end{table}

\subsection{\label{5.1}Static Case}

In the static case we suppose the PH to withdrawal exactly at the
guaranteed rate: $\gamma_{t}=1$.

The static tests 1 and 2 are inspired by \cite{FV}: in their article,
Forsyth and Vetzal price a GLWB contract in a static framework, under
the Black Scholes model with $r=0.04$ and $\sigma=0.15$. The contract
parameters are reported in the Table \ref{tab:CP1}; the contract
type corresponds to\emph{ case 2} in Section \ref{Proj} and \ref{Pric}.

\begin{table}[H]
\begin{centering}
{\small{}}%
\begin{tabular*}{16.5cm}{@{\extracolsep{\fill}}|ll||ll||ll|}
{\small{}Initial age of PH} & {\small{}$65$} & {\small{}Gr. premium} & {\small{}$100$} & {\small{}DB payment} & {\small{}next anniv.}\tabularnewline
{\small{}$G$} & {\small{}$0.05$} & {\small{}Initial fees} & {\small{}$0$} & {\small{}Ratchet} & {\small{}Off/On (annual)}\tabularnewline
{\small{}Withdrawal rate} & {\small{}$1$ per Y} & {\small{}$\alpha_{m}$} & {\small{}$0$} & {\small{}Strategy} & {\small{}static ($\gamma=1$)}\tabularnewline
{\small{}First withdrawal } & {\small{}$1^{st}$ anniv.} & {\small{}Fees taken} & {\small{}cont.ly} &  & \tabularnewline
\end{tabular*}
\par\end{centering}{\small \par}

\caption{\label{tab:CP1}The contract parameters for static tests (except Test
2B).}
\end{table}

They treated two cases: no ratchet, and annual ratchet. In the first
case they get $\alpha_{g}=35.51$ bp and in the second case $\alpha_{g}=64.92$
bp. In Test 1 and Test 2 we introduce respectively stochastic interest
rate and stochastic volatility to analyze the impact of these model
developments on the fair guarantee fee. The parameters for interest
rate and volatility models has been chosen to be plausible.

To compare our results in the Heston model with Kling's ones in \cite{KL}
we performed test 2B. In this case, product parameters are reported
in Table \ref{tab:CP2}, and correspond to \emph{case 1 }in Section
\ref{Proj} and \ref{Pric}.

\begin{table}[H]
\begin{centering}
{\small{}}%
\begin{tabular*}{16.5cm}{@{\extracolsep{\fill}}|ll||ll||ll|}
{\small{}Initial age of PH} & {\small{}$65$} & {\small{}Gr. premium} & {\small{}$100$} & {\small{}DB payment} & {\small{}next anniv. }\tabularnewline
{\small{}$G$} & {\small{}$4.90\%$, $4.19\%$ if ratchet} & {\small{}Initial fees} & {\small{}$4\%$ } & {\small{}Ratchet} & {\small{}Off/On (annual)}\tabularnewline
{\small{}Withdrawal rate} & {\small{}$1$ per Y} & {\small{}$\alpha_{m}$} & {\small{}$151$ bp} & {\small{}Strategy} & {\small{}static ($\gamma=1$)}\tabularnewline
{\small{}First withdrawal } & {\small{}$1^{st}$ anniv.} & {\small{}Fees taken} & {\small{}at the end} &  & \tabularnewline
\end{tabular*}
\par\end{centering}{\small \par}

\caption{\label{tab:CP2}The contract parameters for Test 2B-Static.}
\end{table}

\subsubsection{Test 1-Static: the Black-Scholes Hull-White Model}

In this test we want to price a product according to BS HW model.
We use the same corresponding parameters as in test \cite{FV}. Model
parameters are shown in the Table \ref{tab:mp1s}. Results are available
in Table \ref{tab:Test1A}. 

All four methods behave well and in the configuration D, gave results
consistent with the benchmark. HPDE proved to be the best: all configurations
gave results consistent with the benchmark. Then APDE and SMC, and
HMC gave good results too. SMC performed a little better than HMC:
the first method simulates the underlying value and the interest rate
exactly and so it is enough to simulate the values at each event time.
HMC matches the first three moments of the BS HW $r$ process, but
doesn't reproduce exactly its law: therefore it is right to increase
the number of steps per year. So, for a given run time, we can simulate
less scenarios in HMC than SMC: effectively, the confidence interval
of HMC is larger than SMC's one. Moreover, SMC over performed the
benchmark when using configuration D. Particularly, correlation between
underlying and interest rate has a fundamental role, and it's impact
can be bigger than ratchet's impact: for example, case no ratchet
with $\rho=0.5$ gave a higher price than case ratchet with $\rho=-0.5$
($111$ bp vs $84$ bp).

\subsubsection{Test 2-Static: the Heston Model}

In this test we want to price a product according to the Heston model.
Model parameters are shown in the Table \ref{tab:mp2s}. Results are
shown in Table \ref{tab:Test2A}.

In this Test, MC methods had more problems; PDE methods' values are
close to the benchmark, while MC method's values were far, but compatibles
with the benchmark (BM's value is inside MCs' confidence interval).
Probably, in this case, the benchmark is not very accurate: this is
due to the fact we used SMC to calculate it. If we compare the two
MC approaches, in this case, they both use a third order approximation
and than they become equivalent: HMC proved to be faster than SMC
when using few time steps (we could exploit $+16\%$ simulations in
configuration A), while SMC proved to be slightly faster in high time
steps simulations, because of more time needed to build the volatility
tree ($-3\%$ simulations in configuration D). HPDE showed to be very
stable (case no ratchet, $\rho=-0.5$, $\alpha_{g}$ didn't changed
through configurations B-D), but APDE behaved well to (monotone convergence).
In the Heston model, correlation has a less important role than in
BS HW case: among the different values of $\rho,$ the value of $\alpha_{g}$
changes less then $5$ bp in no-ratchet case, and less than $1.5$
bp in ratchet case.

\subsubsection{Test 2B-Static: the Heston Model}

In this test we want to obtain the results shown in \cite{KL}, where
the contract are priced with MC techniques. The values given in \cite{KL}
are $150$ bp for both cases (no ratchet and ratchet). Model parameters
are given in Table \ref{tab:mp2bs}. Results are available in Table
\ref{tab:Test4}.

In this Test, all methods gave the same results, but not the same
results as in Kling et al. \cite{KL}. One possibility is that we
have misinterpreted some of the contractual specifications in Kling's
paper, leading to some subtle differences in the contracts that we
are considering as compared to theirs, and these discrepancies result
in different fees. Another potential explanation is that a Monte Carlo
method was used to determine the fee by Kling et al.; this may have
introduced a significant error when calculating the fee unless a very
large number of simulations was used. They didn't report a confidence
interval for their results, so it's hard to understand the cause of
the gap. Moreover we can observe that out two MC methods gave larger
confidence intervals than Test 2-Static: probably, the parameters
used for Test 2B-Static shape a harder pricing problem than previous
test, and more simulations should be performed to obtain same quality
results. Also in this case, HPDE proved to be the most stable method.

\subsubsection{Test 3-Static: Hedging }

To reduce financial risks, insurance companies have to hedge the sold
VA: to accomplish this target they must calculate the greeks of products.

In this test we want to show how the different methods can be used
to calculate the main greeks. This can be done through finite differences
for small shocks on the variable. In general, the PDE methods are
ahead w.r.t. MC methods: the price is computed through finite differences
and so the price under shock is already computed. For MC methods this
is quite harder because the pricing has to be repeated changing the
inputs.

To start, we calculate the underlying greek delta, for the products
of Test 1-Static and Test 2-Static. As in this case we don't want
to compute the fair fee $\alpha_{g}$, we fix it arbitrarily. We choose
two values for each model: one for no ratchet case, and one for ratchet
case. The values chosen are such as to cover the costs of the insurer
regardless of the correlation, and may be plausible on a real case.
Results are available in Table \ref{tab:Test3-1} (all values in Table
must be multiplied by $10^{-4}$).

In this Test, we got very accurate results with all method. Anyway,
HPDE proved to be the best: it is the more stable and accurate. We
remark that despite fair fee changes a lot when changing the correlation
parameter $\rho$, the value of Delta changes much less. Delta calculation
proved to be harder in the Heston model case than in the BS HW model
case.

\subsubsection{Test 4-Static: Risk Management }

Mortality and longevity risks are unhedgeable risks. Usually the Risk
Management Team has to calculate the financial reserve taking into
account these risks. Usually extreme scenarios are chosen and policies
are priced according to them. In this test we analyzed how the different
pricing methods behaved under mortality shocks: the mortality probabilities
have been increased by $10\%$ except the last one who's equal to
1. To be brief, we simply report the fair fee for $D$ case. Results
are available in Table \ref{tab:Test4-1}.

In this Test, we got results similar to Test 1-Static and Test 2-Static,
and mortality shocks didn't affect the convergence quality of the
four methods. We observe that mortality shocks reduce the value of
$\alpha_{g}$ (about minus 5 bp) and this means that an increase in
mortality shouldn't be a source of losses for the insurer. Consequently,
insurers should pay attention to longevity risk.

\begin{table}[p]
\begin{centering}
{\small{}}%
\begin{tabular}{c|ccccc|c}
\hline 
{\small{}$S_{0}$} & {\small{}$r$} & {\small{}$curve$} & {\small{}$k$} & {\small{}$\omega$} & {\small{}$\rho$} & {\small{}$\sigma$}\tabularnewline
{\small{}$100$} & {\small{}$0.04$} & {\small{}$flat$} & {\small{}$1.0$} & {\small{}$0.2$} & \emph{\small{}variable} & {\small{}$0.15$}\tabularnewline
\hline 
\end{tabular}
\par\end{centering}{\small \par}

\caption{\label{tab:mp1s}The model parameters about Test 1-Static.}

\vspace{1cm}

\begin{centering}
%
\par\end{centering}

\caption{\label{tab:Test1A}Test 1-Static. In the first Table, the fair fee
for the Black-Scholes Hull-White model, with no ratchet or annual
ratchet. In the second Table the run times for the no-ratchet case
($\rho=-0.5)$. Finally, the plot of relative error (w.r.t. MC value)
for the three methods in the case $\rho=-0.5$ with no ratchet. The
parameters used for this test are available in Table \ref{tab:CP1}
and in Table \ref{tab:mp1s}.}
\end{table}

\begin{table}[p]
\begin{centering}
{\small{}}%
\begin{tabular}{c|ccccc|c}
\hline 
{\small{}$S_{0}$} & {\small{}$V_{0}$} & {\small{}$\theta$} & {\small{}$k$} & {\small{}$\omega$} & {\small{}$\rho$} & {\small{}$r$}\tabularnewline
{\small{}$100$} & {\small{}$0.15^{2}$} & {\small{}$0.15^{2}$} & {\small{}$1.0$} & {\small{}$0.2$} & \emph{\small{}variable} & {\small{}$0.04$}\tabularnewline
\hline 
\end{tabular}
\par\end{centering}{\small \par}

\caption{\label{tab:mp2s}The model parameters about Test 2-Static.}

\vspace{1cm}

\begin{centering}
%
\par\end{centering}

\caption{\label{tab:Test2A} Test 2-Static.In the first Table, the fair fee
for the Heston model, with no ratchet or annual ratchet. In the second
Table the run times for the no-ratchet case ($\rho=-0.5)$. Finally,
the plot of relative error (w.r.t. MC value) for the three methods
in the case $\rho=-0.5$ with no ratchet. The parameters used for
this test are available in Table \ref{tab:CP1} and in Table \ref{tab:mp2s}.}
\end{table}

\begin{table}[p]
\begin{centering}
{\small{}}%
\begin{tabular}{c|ccccc|c}
\hline 
{\small{}$S_{0}$} & {\small{}$V_{0}$} & {\small{}$\theta$} & {\small{}$k$} & {\small{}$\omega$} & {\small{}$\rho$} & {\small{}$r$}\tabularnewline
{\small{}$100$} & {\small{}$0.22^{2}$} & {\small{}$0.22^{2}$} & {\small{}$4.75$} & {\small{}$0.55$} & {\small{}$-0.569$} & {\small{}$0.04$}\tabularnewline
\hline 
\end{tabular}
\par\end{centering}{\small \par}

\caption{\label{tab:mp2bs}The model parameters about Test 2B-Static.}

\vspace{1cm}

\begin{centering}
\begin{tabular}{|>{\centering}p{0.15cm}|>{\centering}p{1.6cm}>{\centering}p{1.6cm}>{\centering}p{0.8cm}>{\centering}p{0.8cm}>{\centering}p{0.7cm}||>{\centering}p{1.6cm}>{\centering}p{1.6cm}>{\centering}p{0.8cm}>{\centering}p{0.8cm}>{\centering}p{0.7cm}|}
\hline 
 & \multicolumn{5}{c|}{\emph{No ratchet}} & \multicolumn{5}{c|}{\emph{Annual ratchet}}\tabularnewline
\cline{2-11} 
 & {\footnotesize{}HMC} & {\footnotesize{}SMC} & {\footnotesize{}HPDE} & {\footnotesize{}APDE} & {\footnotesize{}BM} & {\footnotesize{}HMC} & {\footnotesize{}SMC} & {\footnotesize{}HPDE} & {\footnotesize{}APDE} & {\footnotesize{}BM}\tabularnewline
\hline 
\multicolumn{1}{|c|}{\emph{\scriptsize{}A}} & \textcolor{blue}{\scriptsize{}$138.54\:\pm\:2.70$} & \textcolor{blue}{\scriptsize{}$141.86\:\pm\:2.70$} & \textcolor{blue}{\scriptsize{}$130.83$} & \textcolor{blue}{\scriptsize{}$137.13$} & \multirow{2}{0.7cm}{\textcolor{blue}{\scriptsize{}$131.11$}} & \textcolor{blue}{\scriptsize{}$125.40\:\pm\:2.46$} & \textcolor{blue}{\scriptsize{}$128.33\:\pm\:2.77$} & \textcolor{blue}{\scriptsize{}$117.08$} & \textcolor{blue}{\scriptsize{}$124.19$} & \multirow{2}{0.7cm}{\textcolor{blue}{\scriptsize{}$117.56$}}\tabularnewline
\emph{\scriptsize{}B} & \textcolor{blue}{\scriptsize{}$132.57\:\pm\:1.98$} & \textcolor{blue}{\scriptsize{}$137.16\:\pm\:2.18$} & \textcolor{blue}{\scriptsize{}$130.80$} & \textcolor{blue}{\scriptsize{}$135.76$} &  & \textcolor{blue}{\scriptsize{}$119.33\:\pm\:1.80$} & \textcolor{blue}{\scriptsize{}$123.68\:\pm\:2.00$} & \textcolor{blue}{\scriptsize{}$117.18$} & \textcolor{blue}{\scriptsize{}$122.78$} & \tabularnewline
\hline 
\end{tabular}
\par\end{centering}

\caption{\label{tab:Test4}Test 2B-Static. Fair fee for the Heston model, with
no ratchet or annual ratchet. The parameters used for this test are
available in Table \ref{tab:CP2} and in Table \ref{tab:mp2bs}.}
\end{table}

\begin{table}[p]
\begin{centering}
\begin{tabular}{|c|c>{\centering}p{0.15cm}|ccccc||ccccc|}
\hline 
\multirow{14}{*}{\begin{sideways}
{\small{}Black-Scholes Hull-White}
\end{sideways}} & \multirow{2}{*}{{\small{}$\rho$}} & \multirow{2}{0.15cm}{} & \multicolumn{5}{c|}{\emph{\small{}no ratchet ($\alpha_{g}=150$)}} & \multicolumn{5}{c|}{\emph{\small{}ratchet ($\alpha_{g}=250$)}}\tabularnewline
\cline{4-13} 
 &  &  & {\footnotesize{}HMC} & {\footnotesize{}SMC} & {\footnotesize{}HPDE} & {\footnotesize{}APDE} & {\footnotesize{}MC} & {\footnotesize{}HMC} & {\footnotesize{}SMC} & {\footnotesize{}HPDE} & {\footnotesize{}APDE} & {\footnotesize{}MC}\tabularnewline
\cline{2-13} 
 & \multirow{4}{*}{\begin{sideways}
{\scriptsize{}$-0.5$}
\end{sideways}} & \emph{\scriptsize{}A} & \textcolor{blue}{\scriptsize{}$6055\pm12$} & \textcolor{blue}{\scriptsize{}$6060\pm\phantom{0}8$} & \textcolor{blue}{\scriptsize{}$6058$} & \textcolor{blue}{\scriptsize{}$6034$} & \multirow{2}{*}{\textcolor{blue}{\scriptsize{}$6059$}} & \textcolor{blue}{\scriptsize{}$7123\pm12$} & \textcolor{blue}{\scriptsize{}$7119\pm\phantom{0}8$} & \textcolor{blue}{\scriptsize{}$7118$} & \textcolor{blue}{\scriptsize{}$7128$} & \multirow{2}{*}{\textcolor{blue}{\scriptsize{}$7121$}}\tabularnewline
 &  & \emph{\scriptsize{}B} & \textcolor{blue}{\scriptsize{}$6043\pm\phantom{0}9$} & \textcolor{blue}{\scriptsize{}$6054\pm\phantom{0}4$} & \textcolor{blue}{\scriptsize{}$6058$} & \textcolor{blue}{\scriptsize{}$6050$} &  & \textcolor{blue}{\scriptsize{}$7108\pm\phantom{0}9$} & \textcolor{blue}{\scriptsize{}$7115\pm\phantom{0}4$} & \textcolor{blue}{\scriptsize{}$7119$} & \textcolor{blue}{\scriptsize{}$7126$} & \tabularnewline
\hline 
\end{tabular}
\par\end{centering}

\vspace{0.2cm}

\begin{centering}
\begin{tabular}{|c|c>{\centering}p{0.15cm}|ccccc||ccccc|}
\hline 
\multirow{14}{*}{\begin{sideways}
{\small{}Heston}
\end{sideways}} & \multirow{2}{*}{{\small{}$\rho$}} & \multirow{2}{0.15cm}{} & \multicolumn{5}{c|}{\emph{\small{}no ratchet ($\alpha_{g}=50$)}} & \multicolumn{5}{c|}{\emph{\small{}ratchet ($\alpha_{g}=100$)}}\tabularnewline
\cline{4-13} 
 &  &  & {\footnotesize{}HMC} & {\footnotesize{}SMC} & {\footnotesize{}HPDE} & {\footnotesize{}APDE} & {\footnotesize{}MC} & {\footnotesize{}HMC} & {\footnotesize{}SMC} & {\footnotesize{}HPDE} & {\footnotesize{}APDE} & {\footnotesize{}MC}\tabularnewline
\cline{2-13} 
 & \multirow{4}{*}{\begin{sideways}
{\scriptsize{}$-0.5$}
\end{sideways}} & \emph{\scriptsize{}A} & \textcolor{blue}{\scriptsize{}$7870\pm20$} & \textcolor{blue}{\scriptsize{}$7856\pm23$} & \textcolor{blue}{\scriptsize{}$7875$} & \textcolor{blue}{\scriptsize{}$7867$} & \multirow{2}{*}{\textcolor{blue}{\scriptsize{}$7875$}} & \textcolor{blue}{\scriptsize{}$8509\pm14$} & \textcolor{blue}{\scriptsize{}$8499\pm15$} & \textcolor{blue}{\scriptsize{}$8502$} & \textcolor{blue}{\scriptsize{}$8512$} & \multirow{2}{*}{\textcolor{blue}{\scriptsize{}$8509$}}\tabularnewline
 &  & \emph{\scriptsize{}B} & \textcolor{blue}{\scriptsize{}$7873\pm15$} & \textcolor{blue}{\scriptsize{}$7868\pm16$} & \textcolor{blue}{\scriptsize{}$7875$} & \textcolor{blue}{\scriptsize{}$7873$} &  & \textcolor{blue}{\scriptsize{}$8506\pm10$} & \textcolor{blue}{\scriptsize{}$8503\pm11$} & \textcolor{blue}{\scriptsize{}$8506$} & \textcolor{blue}{\scriptsize{}$8516$} & \tabularnewline
\hline 
\end{tabular}
\par\end{centering}

\caption{\label{tab:Test3-1}Test 3-Static. Delta calculation for the Black-Scholes
Hull-White model and the Heston model, with no ratchet or annual ratchet
(these value must be multiplied by $10^{-4}$). The parameters used
for this test are available in Table \ref{tab:CP1}, in Table \ref{tab:mp1s}
and in Table \ref{tab:mp2s}.}
\end{table}

\begin{table}[p]
\begin{centering}
\begin{tabular}{|c|c|r@{\extracolsep{0pt}.}lr@{\extracolsep{0pt}.}lr@{\extracolsep{0pt}.}lr@{\extracolsep{0pt}.}lr@{\extracolsep{0pt}.}l||r@{\extracolsep{0pt}.}lr@{\extracolsep{0pt}.}lr@{\extracolsep{0pt}.}lr@{\extracolsep{0pt}.}lr@{\extracolsep{0pt}.}l|}
\hline 
\multicolumn{2}{|c|}{} & \multicolumn{10}{c|}{\emph{\small{}No ratchet}} & \multicolumn{10}{c|}{\emph{\small{}Ratchet}}\tabularnewline
\cline{3-22} 
\multicolumn{1}{|c}{} &  & \multicolumn{2}{c}{{\footnotesize{}HMC}} & \multicolumn{2}{c}{{\footnotesize{}SMC}} & \multicolumn{2}{c}{{\footnotesize{}HPDE}} & \multicolumn{2}{c}{{\footnotesize{}APDE}} & \multicolumn{2}{c||}{{\footnotesize{}BM}} & \multicolumn{2}{c}{{\footnotesize{}HMC}} & \multicolumn{2}{c}{{\footnotesize{}SMC}} & \multicolumn{2}{c}{{\footnotesize{}HPDE}} & \multicolumn{2}{c}{{\footnotesize{}APDE}} & \multicolumn{2}{c|}{{\footnotesize{}BM}}\tabularnewline
\hline 
\multirow{6}{*}{\begin{sideways}
{\footnotesize{}BS HW}
\end{sideways}} & \multirow{2}{*}{\textcolor{blue}{\emph{\scriptsize{}$-0.5$}}} & \multicolumn{2}{c}{\textcolor{blue}{\scriptsize{}$41.89$}} & \multicolumn{2}{c}{\textcolor{blue}{\scriptsize{}$41.75$}} & \multicolumn{2}{c}{\textcolor{blue}{\scriptsize{}$41.75$}} & \multicolumn{2}{c}{\textcolor{blue}{\scriptsize{}$41.85$}} & \multicolumn{2}{c||}{\textcolor{blue}{\scriptsize{}$41.84$}} & \multicolumn{2}{c}{\textcolor{blue}{\scriptsize{}$75.96$}} & \multicolumn{2}{c}{\textcolor{blue}{\scriptsize{}$75.91$}} & \multicolumn{2}{c}{\textcolor{blue}{\scriptsize{}$75.92$}} & \multicolumn{2}{c}{\textcolor{blue}{\scriptsize{}$76.20$}} & \multicolumn{2}{c|}{\textcolor{blue}{\scriptsize{}$76.00$}}\tabularnewline
\hline 
\end{tabular}
\par\end{centering}

\vspace{0.3cm}

\begin{centering}
\begin{tabular}{|c|c|r@{\extracolsep{0pt}.}lr@{\extracolsep{0pt}.}lr@{\extracolsep{0pt}.}lr@{\extracolsep{0pt}.}lr@{\extracolsep{0pt}.}l||r@{\extracolsep{0pt}.}lr@{\extracolsep{0pt}.}lr@{\extracolsep{0pt}.}lr@{\extracolsep{0pt}.}lr@{\extracolsep{0pt}.}l|}
\hline 
\multicolumn{2}{|c|}{} & \multicolumn{10}{c|}{\emph{\small{}No ratchet}} & \multicolumn{10}{c|}{\emph{\small{}Ratchet}}\tabularnewline
\cline{3-22} 
\multicolumn{1}{|c}{} &  & \multicolumn{2}{c}{{\footnotesize{}HMC}} & \multicolumn{2}{c}{{\footnotesize{}SMC}} & \multicolumn{2}{c}{{\footnotesize{}HPDE}} & \multicolumn{2}{c}{{\footnotesize{}APDE}} & \multicolumn{2}{c||}{{\footnotesize{}BM}} & \multicolumn{2}{c}{{\footnotesize{}HMC}} & \multicolumn{2}{c}{{\footnotesize{}SMC}} & \multicolumn{2}{c}{{\footnotesize{}HPDE}} & \multicolumn{2}{c}{{\footnotesize{}APDE}} & \multicolumn{2}{c|}{{\footnotesize{}BM}}\tabularnewline
\hline 
\multirow{6}{*}{\begin{sideways}
{\footnotesize{}Heston}
\end{sideways}} & \multicolumn{1}{c|}{\emph{\footnotesize{}$-0.5$}} & \multicolumn{2}{c}{\textcolor{blue}{\scriptsize{}$34.48$}} & \multicolumn{2}{c}{\textcolor{blue}{\scriptsize{}$34.19$}} & \multicolumn{2}{c}{\textcolor{blue}{\scriptsize{}$33.88$}} & \multicolumn{2}{c}{\textcolor{blue}{\scriptsize{}$33.92$}} & \multicolumn{2}{c||}{\textcolor{blue}{\scriptsize{}$33.88$}} & \multicolumn{2}{c}{\textcolor{blue}{\scriptsize{}$56.03$}} & \multicolumn{2}{c}{\textcolor{blue}{\scriptsize{}$56.83$}} & \multicolumn{2}{c}{\textcolor{blue}{\scriptsize{}$55.54$}} & \multicolumn{2}{c}{\textcolor{blue}{\scriptsize{}$55.61$}} & \multicolumn{2}{c|}{\textcolor{blue}{\scriptsize{}$55.56$}}\tabularnewline
\hline 
\end{tabular}
\par\end{centering}

\caption{\label{tab:Test4-1}Test 4-Static. Impact of $+10\%$ mortality shocks
of fair fee. The parameters used for this test are available in Table
\ref{tab:CP1}, in Table \ref{tab:mp1s} and in Table \ref{tab:mp2s}.}
\end{table}

\FloatBarrier

\subsection{\label{5.2}Dynamic case}

In the Dynamic case, the policy holder is supposed to choose the worst
strategy from an hedger point of view, changing the value of $\gamma_{t}$.
The PH can withdraw more ($1\leq\gamma_{t}\leq2$) or less ($0\leq\gamma_{t}\leq1$)
than the standard rate (see \ref{gamma} for more details).

In this pricing framework, we refer to the prices in Forsyth and Vetzal
\cite{FV}: in their article, the authors price a GLWB contract in
a static framework, under the Black Scholes model with $r=0.04$ and
$\sigma=0.15$. The contract parameters are reported in the Table
\ref{tab:CP3} (Table 6.7 in \cite{FV}). They treated two cases:
no ratchet, and ratchet every 3 years; both of them corresponds to
\emph{case 4} in Section \ref{Proj} and \ref{Pric}. In the first
case they got $\alpha_{g}=63.1$ bp and in the second case $\alpha_{g}=70.7$
bp. In Test 1 and 2 we introduce respectively stochastic interest
rate and stochastic volatility to analyze the impact of these model
developments on the fair guarantee fee. The parameters for interest
rate and volatility models are the same as the static case. 

\begin{table}
\begin{centering}
{\small{}}%
\begin{tabular}{|ll||ll||ll||ll|}
{\small{}Initial age of PH} & {\small{}$65$} & {\small{}Gr. premium} & {\small{}$100$} & {\small{}DB payment} & {\small{}cont.ly} & {\small{}Strategy} & {\small{}Dynamic}\tabularnewline
{\small{}$G$} & {\small{}$0.05$} & {\small{}Initial fees} & {\small{}$0$} & {\small{}Ratchet} & {\small{}Off/On } & {\small{}Bonus } & {\small{}$5\%$}\tabularnewline
{\small{}Withdrawal rate} & {\small{}$1$ per Y} & {\small{}$\alpha_{m}$} & {\small{}$0$} & {\small{}Ratchet rate} & {\small{}every 3 Ys} & {\small{}$\kappa\left(t\right)$} & {\small{}see tab below}\tabularnewline
{\small{}First withdrawal } & {\small{}$1^{st}$ anniv.} & {\small{}Fees taken} & {\small{}cont.ly} &  &  &  & \tabularnewline
\end{tabular}
\par\end{centering}{\small \par}

\vspace{0.2cm}

\begin{centering}
{\small{}}%
\begin{tabular}{|c|c|c|c|c|c|c|}
\multirow{2}{*}{{\small{}$\kappa\left(t\right)$}} & {\small{}$0\leq t\leq1$} & {\small{}$1<t\leq2$} & {\small{}$2<t\leq3$} & {\small{}$3<t\leq4$} & {\small{}$4<t\leq5$} & {\small{}$t>5$}\tabularnewline
 & {\small{}$5\%$} & {\small{}$4\%$} & {\small{}$3\%$} & {\small{}$2\%$} & {\small{}$1\%$} & {\small{}$0\%$}\tabularnewline
\end{tabular}
\par\end{centering}{\small \par}

\caption{\label{tab:CP3}The contract parameters used in the dynamic case.}
\end{table}

Here a brief summary of the numerical results for this Section.

\subsubsection{Test 1-Dynamic: the Black-Scholes Hull-White Model}

Test 1-Dynamic is the dynamic case of Test 1-Static. Model parameters
are shown in Table \ref{tab:mp1s}. Results are available in Table
\ref{tab:Test1A-1}.

In this Test, PDE methods proved to be more efficient than MC ones.
In fact MC ones use Longstaff-Schwartz method to find the optimal
withdrawal: this method needs a lot of scenarios to approximate through
the least squares approach the value of the police for a given set
of variable, and the regression is time demanding. Then, working at
fixed time, we could perform fewer scenarios than static case (around
$10\%$), while PDE methods used almost the same parameters as in
static case. Moreover the regression problem proved to be be hard:
sometimes, excluding the value $\gamma=0$ among the possible values
that the PH can chose (therefore excluding no withdrawal case), we
got higher values for $\alpha_{g}$. This means that the regression
isn't very accurate, and sometimes we fail to find the optimal withdrawal:
that's why, using MC methods we usually find smaller value for $\alpha_{g}$
than the right value. In particular, we excluded the value $\gamma=0$
while using configurations A and B. We would remark that also the
benchmark is affected by these computation problems and in case no-ratchet
with $\rho=-0.5$ we got a small value for benchmark then PDE method
(around $261$ vs $266$). Another thing to remark is that MC methods
behaved better while ratchets were considered: maybe in this case
in it easier to find the best strategy. The two MC methods proved
to be equivalent: the differences in scenario generation's run-time
are negligible because most of the time is spent in finding the best
withdrawal. Both APDE and HPDE method gave good and stable results,
but HPDE performed better in case A.

\subsubsection{Test 2-Dynamic: the Heston Model}

Test 2-Dynamic is the dynamic case of Test 2-Static. Model parameters
are shown in Table \ref{tab:mp2s}. Results are available in Table
\ref{tab:Test2A-1}.

In this Test, things are similar to Test 1-Dynamic, but the optimization
problem seemed to be easier than in Test 1-Dynamic: MC methods converged
better, especially when using high level configurations. PDE methods
behaved good as usual, and HPDE method proved to be a bit better then
APDE method. The two MC methods proved to be equivalent. We note that,
in Heston model case, Dynamic strategy increase the value of $\alpha_{g}$
less than in BS HW case: probably, playing on interest rate, let the
PH to gain more than playing on volatility.

\subsubsection{Test 3-Dynamic: Hedging}

Test 3-Dynamic is the dynamic case of Test 3-Static. Results are available
in Table \ref{tab:Test3-1-1}.

In this Test, we got good results with all methods, but MC methods
proved to be inaccurate while using configurations A and B. The range
of possible values for Delta increased with regard to Test 3-Static.
The two MC methods proved to be equivalent.

\subsubsection{Test 4-Dynamic: Risk Management}

Test 4-Dynamic is the dynamic case of Test 4-Static. Results are available
in Table \ref{tab:Test4-1-1}.

In this Test, we got similar results with regard to Test 4-Static:
the fees reduced a little (around $20$ bp in the BS HW model case
and around $6$ bp in the Heston model case).

\begin{figure}[p]
\begin{centering}
\includegraphics[scale=0.8]{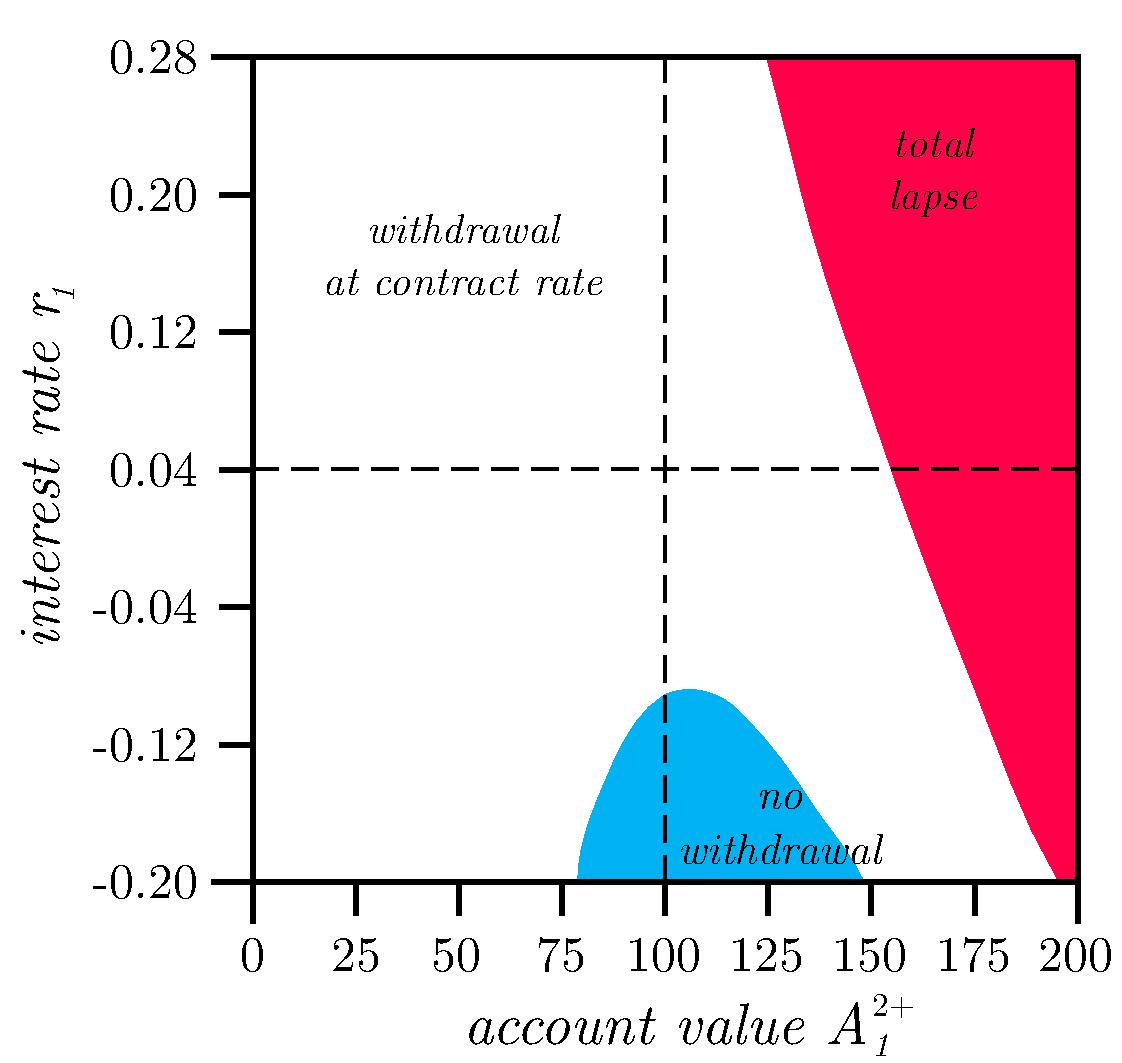}\hspace{1 cm}\includegraphics[scale=0.8]{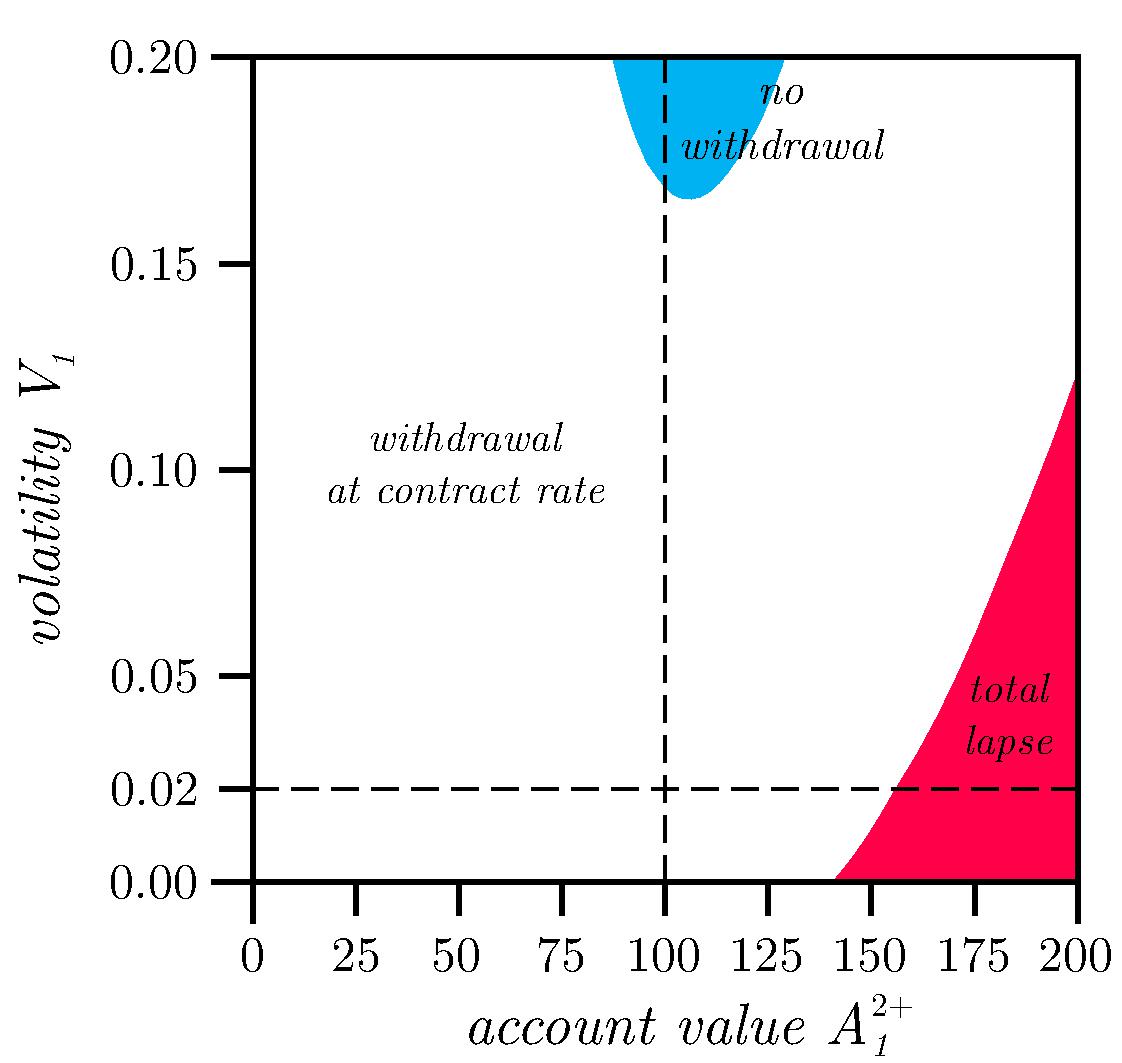}
\par\end{centering}

\caption{\label{fig:Optimal-strategy}Optimal strategy at the first event time
$\left(t=1\right)$ for the BS HW model and the Heston model, assuming
$B_{1}^{2+}=100$. Model parameters are available in Tables \ref{tab:mp1s}
and \ref{tab:mp2s}. Product parameters are available in Table \ref{tab:CP3},
and $\alpha_{g}=135$ bp for both cases.}
\end{figure}

In Figure \ref{fig:Optimal-strategy}, we present, as an example,
the optimal strategy at time $t=1$ in two different cases. We can
see the best strategy at time $t=1$. We can see how it is worth to
lapse when the account value reaches high values, and especially when
interest rate is high or volatility low. It's more difficult to understand
when do no withdrawal: there must be a convenient mix of all the variables.

\begin{table}[p]
\begin{centering}

\par\end{centering}

\caption{\label{tab:Test1A-1}Test 1-Dynamic. In the first Table, the fair
fee for the Black-Scholes Hull-White model, with no ratchet or annual
ratchet. In the second Table the run times for the no-ratchet case
($\rho=-0.5)$. Finally, the plot of relative error (w.r.t. MC value)
for the three methods in the case $\rho=-0.5$ with no ratchet. The
parameters used for this test are available in Table \ref{tab:CP3}
and in Table \ref{tab:mp1s}.}
\end{table}

\begin{table}[p]
\begin{centering}
%
\par\end{centering}

\caption{\label{tab:Test2A-1}Test 2-Dynamic. In the first Table, the fair
fee for the Heston model, with no ratchet or annual ratchet. In the
second Table the run times for the no-ratchet case ($\rho=-0.5)$.
Finally, the plot of relative error (w.r.t. MC value) for the three
methods in the case $\rho=-0.5$ with no ratchet. The parameters used
for this test are available in Table \ref{tab:CP3} and in Table \ref{tab:mp2s}.}
\end{table}

\begin{table}[p]
\begin{centering}
{\small{}}%
\begin{tabular}{|c|cc|ccccc||ccccc|}
\hline 
\multirow{14}{*}{\begin{sideways}
{\small{}Black-Scholes Hull-White}
\end{sideways}} & \multirow{2}{*}{{\small{}$\rho$}} & \multirow{2}{*}{} & \multicolumn{5}{c|}{\emph{\small{}no ratchet ($\alpha_{g}=300$)}} & \multicolumn{5}{c}{\emph{\small{}ratchet ($\alpha_{g}=350$)}}\tabularnewline
\cline{4-13} 
 &  &  & {\footnotesize{}HMC} & {\footnotesize{}SMC} & {\footnotesize{}HPDE} & {\footnotesize{}APDE} & {\footnotesize{}BM} & {\footnotesize{}HMC} & {\footnotesize{}SMC} & {\footnotesize{}HPDE} & {\footnotesize{}APDE} & {\footnotesize{}BM}\tabularnewline
\cline{2-13} 
 & \multirow{4}{*}{\begin{sideways}
{\scriptsize{}$-0.5$}
\end{sideways}} & \emph{\scriptsize{}A} & \textcolor{blue}{\scriptsize{}$8513\pm573$} & \textcolor{blue}{\scriptsize{}$7951\pm557$} & \textcolor{blue}{\scriptsize{}$8078$} & \textcolor{blue}{\scriptsize{}$8097$} & \multirow{2}{*}{\textcolor{blue}{\scriptsize{}$8078$}} & \textcolor{blue}{\scriptsize{}$8959\pm542$} & \textcolor{blue}{\scriptsize{}$8279\pm551$} & \textcolor{blue}{\scriptsize{}$8148$} & \textcolor{blue}{\scriptsize{}$8200$} & \multirow{2}{*}{\textcolor{blue}{\scriptsize{}$8157$}}\tabularnewline
 &  & \emph{\scriptsize{}B} & \textcolor{blue}{\scriptsize{}$8220\pm347$} & \textcolor{blue}{\scriptsize{}$8304\pm305$} & \textcolor{blue}{\scriptsize{}$8081$} & \textcolor{blue}{\scriptsize{}$8091$} &  & \textcolor{blue}{\scriptsize{}$7922\pm324$} & \textcolor{blue}{\scriptsize{}$8104\pm330$} & \textcolor{blue}{\scriptsize{}$8151$} & \textcolor{blue}{\scriptsize{}$8180$} & \tabularnewline
\hline 
\end{tabular}
\par\end{centering}{\small \par}

\vspace{0.2cm}

\begin{centering}
\begin{tabular}{|c|cc|ccccc||ccccc|}
\hline 
\multirow{14}{*}{\begin{sideways}
{\small{}Heston}
\end{sideways}} & \multirow{2}{*}{{\small{}$\rho$}} & \multirow{2}{*}{} & \multicolumn{5}{c|}{\emph{\small{}no ratchet ($\alpha_{g}=75$)}} & \multicolumn{5}{c}{\emph{\small{}ratchet ($\alpha_{g}=100$)}}\tabularnewline
\cline{4-13} 
 &  &  & {\footnotesize{}HMC} & {\footnotesize{}SMC} & {\footnotesize{}HPDE} & {\footnotesize{}APDE} & {\footnotesize{}BM} & {\footnotesize{}HMC} & {\footnotesize{}SMC} & {\footnotesize{}HPDE} & {\footnotesize{}APDE} & {\footnotesize{}BM}\tabularnewline
\cline{2-13} 
 & \multirow{4}{*}{\begin{sideways}
{\scriptsize{}$-0.5$}
\end{sideways}} & \emph{\scriptsize{}A} & \textcolor{blue}{\scriptsize{}$8181\pm472$} & \textcolor{blue}{\scriptsize{}$7794\pm524$} & \textcolor{blue}{\scriptsize{}$8436$} & \textcolor{blue}{\scriptsize{}$8429$} & \multirow{2}{*}{\textcolor{blue}{\scriptsize{}$8432$}} & \textcolor{blue}{\scriptsize{}$8349\pm293$} & \textcolor{blue}{\scriptsize{}$8374\pm324$} & \textcolor{blue}{\scriptsize{}$8477$} & \textcolor{blue}{\scriptsize{}$8474$} & \multirow{2}{*}{\textcolor{blue}{\scriptsize{}$8481$}}\tabularnewline
 &  & \emph{\scriptsize{}B} & \textcolor{blue}{\scriptsize{}$8440\pm250$} & \textcolor{blue}{\scriptsize{}$8383\pm223$} & \textcolor{blue}{\scriptsize{}$8436$} & \textcolor{blue}{\scriptsize{}$8436$} &  & \textcolor{blue}{\scriptsize{}$8304\pm174$} & \textcolor{blue}{\scriptsize{}$8527\pm154$} & \textcolor{blue}{\scriptsize{}$8479$} & \textcolor{blue}{\scriptsize{}$8480$} & \tabularnewline
\hline 
\end{tabular}
\par\end{centering}

\caption{\label{tab:Test3-1-1}Test 3-Dynamic. Delta calculation for the Black-Scholes
Hull-White model and the Heston model, with no ratchet or ratchet
(once every 3 years); these value must be multiplied by $10^{-4}$.
The parameters used for this test are available in Table \ref{tab:CP3},
in Table \ref{tab:mp1s} and in Table \ref{tab:mp2s}.}
\end{table}

\begin{table}[p]
\begin{centering}
\begin{tabular}{|c|c|r@{\extracolsep{0pt}.}lr@{\extracolsep{0pt}.}lr@{\extracolsep{0pt}.}lr@{\extracolsep{0pt}.}lr@{\extracolsep{0pt}.}l||r@{\extracolsep{0pt}.}lr@{\extracolsep{0pt}.}lr@{\extracolsep{0pt}.}lr@{\extracolsep{0pt}.}lr@{\extracolsep{0pt}.}l|}
\hline 
\multicolumn{2}{|c|}{} & \multicolumn{10}{c|}{\emph{\small{}No ratchet}} & \multicolumn{10}{c}{\emph{\small{}Ratchet}}\tabularnewline
\cline{3-22} 
\multicolumn{1}{|c}{} &  & \multicolumn{2}{c}{{\footnotesize{}HMC}} & \multicolumn{2}{c}{{\footnotesize{}SMC}} & \multicolumn{2}{c}{{\footnotesize{}HPDE}} & \multicolumn{2}{c}{{\footnotesize{}APDE}} & \multicolumn{2}{c||}{{\footnotesize{}BM}} & \multicolumn{2}{c}{{\footnotesize{}HMC}} & \multicolumn{2}{c}{{\footnotesize{}SMC}} & \multicolumn{2}{c}{{\footnotesize{}HPDE}} & \multicolumn{2}{c}{{\footnotesize{}APDE}} & \multicolumn{2}{c|}{{\footnotesize{}BM}}\tabularnewline
\hline 
\multirow{6}{*}{\begin{sideways}
{\scriptsize{}BS HW}
\end{sideways}} & \multirow{2}{*}{\emph{\scriptsize{}$-0.5$}} & \multicolumn{2}{c}{\textcolor{blue}{\scriptsize{}$76.64$}} & \multicolumn{2}{c}{\textcolor{blue}{\scriptsize{}$75.74$}} & \multicolumn{2}{c}{\textcolor{blue}{\scriptsize{}$76.85$}} & \multicolumn{2}{c}{\textcolor{blue}{\scriptsize{}$76.19$}} & \multicolumn{2}{c||}{\textcolor{blue}{\scriptsize{}$76.16$}} & \multicolumn{2}{c}{\textcolor{blue}{\scriptsize{}$85.10$}} & \multicolumn{2}{c}{\textcolor{blue}{\scriptsize{}$85.47$}} & \multicolumn{2}{c}{\textcolor{blue}{\scriptsize{}$86.59$}} & \multicolumn{2}{c}{\textcolor{blue}{\scriptsize{}$85.84$}} & \multicolumn{2}{c|}{\textcolor{blue}{\scriptsize{}$86.97$}}\tabularnewline
\hline 
\end{tabular}
\par\end{centering}

\vspace{1cm}

\begin{centering}
\begin{tabular}{|c|c|r@{\extracolsep{0pt}.}lr@{\extracolsep{0pt}.}lr@{\extracolsep{0pt}.}lr@{\extracolsep{0pt}.}lr@{\extracolsep{0pt}.}l||r@{\extracolsep{0pt}.}lr@{\extracolsep{0pt}.}lr@{\extracolsep{0pt}.}lr@{\extracolsep{0pt}.}lr@{\extracolsep{0pt}.}l|}
\hline 
\multicolumn{2}{|c|}{} & \multicolumn{10}{c|}{\emph{\small{}No ratchet}} & \multicolumn{10}{c}{\emph{\small{}Ratchet}}\tabularnewline
\cline{3-22} 
\multicolumn{1}{|c}{} &  & \multicolumn{2}{c}{{\footnotesize{}HMC}} & \multicolumn{2}{c}{{\footnotesize{}SMC}} & \multicolumn{2}{c}{{\footnotesize{}HPDE}} & \multicolumn{2}{c}{{\footnotesize{}APDE}} & \multicolumn{2}{c||}{{\footnotesize{}BM}} & \multicolumn{2}{c}{{\footnotesize{}HMC}} & \multicolumn{2}{c}{{\footnotesize{}SMC}} & \multicolumn{2}{c}{{\footnotesize{}HPDE}} & \multicolumn{2}{c}{{\footnotesize{}APDE}} & \multicolumn{2}{c|}{{\footnotesize{}BM}}\tabularnewline
\hline 
\multirow{6}{*}{\begin{sideways}
{\scriptsize{}Heston}
\end{sideways}} & \multirow{2}{*}{\emph{\scriptsize{}$-0.5$}} & \multicolumn{2}{c}{\textcolor{blue}{\scriptsize{}$57.74$}} & \multicolumn{2}{c}{\textcolor{blue}{\scriptsize{}$58.77$}} & \multicolumn{2}{c}{\textcolor{blue}{\scriptsize{}$58.78$}} & \multicolumn{2}{c}{\textcolor{blue}{\scriptsize{}$58.32$}} & \multicolumn{2}{c||}{\textcolor{blue}{\scriptsize{}$58.56$}} & \multicolumn{2}{c}{\textcolor{blue}{\scriptsize{}$63.73$}} & \multicolumn{2}{c}{\textcolor{blue}{\scriptsize{}$64.50$}} & \multicolumn{2}{c}{\textcolor{blue}{\scriptsize{}$64.74$}} & \multicolumn{2}{c}{\textcolor{blue}{\scriptsize{}$64.22$}} & \multicolumn{2}{c|}{\textcolor{blue}{\scriptsize{}$64.59$}}\tabularnewline
\hline 
\end{tabular}
\par\end{centering}

\caption{\label{tab:Test4-1-1}Test 4-Dynamic. Impact of $+10\%$ mortality
shocks of fair fee. The parameters used for this test are available
in Table \ref{tab:CP3}, in Table \ref{tab:mp1s} and in Table \ref{tab:mp2s}.}
\end{table}

\clearpage

\section{\label{6}Conclusions}

In this article we have developed four methods to price GLWB contracts
under different conditions. Regarding the stochastic model, both stochastic
interest rates and stochastic volatility effects have been considered.
Regarding the policy holder behavior, both static and dynamic strategy
have been considered.

Since GLWB variable annuities are such a long maturity products, the
effects of stochastic interest rates and stochastic volatility cannot
be overlook. In particular, the impact of stochastic rate seems to
be more relevant. Also Forsyth and Vetzal in \cite{FV} used regime
switching model having both stochastic interest rate and volatility,
but our approach, based on SDE, is more realistic, and suitable for
hedging.

All four methods gave compatible results both for pricing and delta
calculation. The fair hedging fee (i.e. the cost of maintaining the
replicating portfolio) is determined using a sequence of parameters'
refinements. The PDE methods proved to be not very expensive, while
MC methods proved to be more expensive. The Hybrid PDE seemed to be
the more performing than the others for its convergence speed and
stability of results. Also ADI PDE behaved very well but the implementation
was harder then Hybrid PDE one. In the BS HW model case, Standard
MC thanks to its exact simulation outperformed the Hybrid Method while,
in the Heston model case, the MC methods proved to be roughly equivalent,
even if the Hybrid MC was easier to be implemented.

As we said before, PDE methods proved to be much more efficient than
MC methods, especially in Dynamic case where is much more simple to
implement the optimal withdrawal choice. Similarity reduction reduces
the problem dimension to a 2D problem and therefore PDE methods perform
well. Anyway, we have to remark that MC methods offer a confidence
interval for the result, they are useful in risk measures calculation
(for example VAR or ES), and they are preferred by insurance companies
because of their attachment to the idea of scenario. 

A future development that could be treated is to combine stochastic
interest rate and stochastic volatility: the combined model could
be an element of greater realism.

We conclude by pointing out that our methods are quite flexible in
that they can accommodate a wide variety of policy holder withdrawal
strategies such as ones derived from utility-based models.


\newpage \clearpage
\begin{thebibliography}{9}
\small{
\bibitem{AA}
\textsc{A. Alfonsi} (2010). High order discretization schemes for the CIR process: application to Affine Term Structure and Heston models. \textit{Mathematics of Computation}, Vol. 79, No. 269, pp. 209-237. 

\bibitem{AC}
\textsc{E. Appolloni, L. Caramellino A. Zanette} 2014. A robust tree method for pricing American options with the Cox-Ingersoll-Ross interest rate model. \textit{IMA J Management Math} first published online January 15, 2014 doi:10.1093/imaman/dpt030.

\bibitem{bips}
\textsc{A. R. Bacinello, P. Millossovich, A. Olivieri, E. Pitacco} (2011). Variable annuities: A unifying valuation approach. \textit{Insurance: Mathematics and Economics} 49, pp. 285-297.

\bibitem{BCZ0}
\textsc{M. Briani, L. Caramellino, A. Zanette} (2014). A hybrid tree-finite difference approach for the Heston model. Preprint, arXiv:1307.7178v2.

\bibitem{BCZ}
\textsc{M. Briani, L. Caramellino, A. Zanette} (2015). Numerical approximations for Heston-Hull-White type models. Preprint, arXiv:1503.03705 .

\bibitem{BF}
\textsc{A. Belanger, P. Forsyth,  G. Labahn} (2009). Valuing the guaranteed minimum death benet clause with partial withdrawals. \textit{Applied Mathematical Finance} 16, pp. 451-496.

\bibitem{BM}
\textsc{D. Brigo, F. Mercurio} (2006). Interest rate models-Theory and practice. \textit{Springer}, Berlin.

\bibitem{CF}
\textsc{Z. Chen, K. Vetzal, P. Forsyth} (2008). The effect of modelling parameters on the value of GMWB guarantees. \textit{Insurance: Mathematics and Economics} 43, pp. 165-173.

\bibitem{FV}
\textsc{P. Forsyth, K.Vetzal} (2014). An optimal stochastic control framework for determining the cost of hedging of variable annuities. \textit{Journal of Economic Dynamics and Control} 44 (2014), pp. 29-53.

\bibitem{GL}
\textsc{P. Gaillardetz,   J. Lakhmiri} (2011). A new premium principle for equity indexed annuities. \textit{Journal of Risk and Insurance} 78, 245-265.

\bibitem{He}
\textsc{S. Heston} (1993): A closed-form solution for options with stochastic volatility with applications to bond and currency options. \textit{The Review of Financial Studies}, Vol. 6, No. 2, pp. 327-343.

\bibitem{HK}
\textsc{D. Holz, A. Kling, J. Ru\ss} (2007). GMWB for life: an analysis of lifelong withdrawal guarantees. Working paper.


\bibitem{HH}
\textsc{T. Haentjens, K. J. In 't Hout} (2012): Alternating direction implicit finite difference schemes for the Heston-Hull-WHite partial differential equation. The Journal of Computation Finance (83-110), Vol. 16, No. 1, Fall 2012.

\bibitem{HW}
\textsc{J. Hull, A. White} (1994). Numerical procedures for implementing term structure models I: single factor models. \textit{The Journal of Derivatives Fall}, 716.

\bibitem{KL}
\textsc{A. Kling, F. Ruez, J. Ru\ss} (2014).  The impact of stochastic volatility on pricing, hedging, and hedge efficiency of variable annuity guarantees. \textit{European Actuarial Journal}, Vol. 4, No. 2, pp. 281-314.

\bibitem{LS}
\textsc{F. A. Longstaff, E. S. Schwartz} (2001). Valuing american options by simulation: a simple least-squares approach. \textit{The Review of Financial Studies} Spring 2001, Vol. 14, No. 1, pp. 113-147.

\bibitem{MS}
\textsc{M. A. Milevsky,  T. S. Salisbury} (2006). Financial valuation of guaranteed minimum withdrawal benets. \textit{Insurance: Mathematics and Economics} 38, pp. 21-38.

\bibitem{NR}
\textsc{D. B. Nelson, K. Ramaswamy} (1990). Simple binomial processes as diffusion approximations in financial models. \textit{The Review of Financial Studies} 1990, Vol. 3, No. 3, pp. 393-430.

\bibitem{SB}
\textsc{P. Shah, D. Bertsimas} (2008). An analysis of the guaranteed withdrawal benets for life option. Working paper, Sloan School of Management, MIT.

\bibitem{VO}
\textsc{V. Ostrovski}  (2013). Efficient and exact simulation of the Hull-White model. Available at SSRN: http://ssrn.com/abstract=2304848 or http://dx.doi.org/10.2139/ssrn.2304848.
}
\end{thebibliography}
\end{document}